\documentclass[12pt]{article}
\usepackage{graphicx}
\usepackage{amsmath, amssymb, amsthm}
\usepackage{booktabs}
\usepackage{float}
\usepackage{longtable}
\usepackage{geometry}
\usepackage{natbib}
\usepackage{mathrsfs}
\usepackage{multirow}
\usepackage{comment}
\usepackage{setspace}
\usepackage{bbm}
\usepackage{bm}
\usepackage{amsfonts}
\usepackage{xcolor}
\usepackage{pslatex}
\usepackage{setspace}
\usepackage[utf8]{inputenc}
\usepackage{bbm}
\usepackage{caption}
\usepackage{subcaption}
\usepackage{graphics,epsfig}
\usepackage{hyperref}
\usepackage{color}
\usepackage[utf8]{inputenc}
\usepackage{pgfplots}
\pgfplotsset{compat=1.17}
\usepgfplotslibrary{statistics}

\title{\bfseries Genomic Influence of a Key Transcription Factor in Male Glandular Malignancy}

  \author{
    Allison Powell$^{1}$ \and Paramahansa Pramanik$^{2,3}$
}
 
  \date{\small
    $^{1}$School of Computing, University of South Alabama, Mobile, AL 36688, United States.\\
    \texttt{ahp2221@jagmail.southalabama.edu} \\
    \vspace{0.5em}
    $^{2}$Department of Mathematics and Statistics, University of South Alabama, Mobile, AL 36688, United States.\\
    $^{3}$Corresponding author, \texttt{ppramanik@southalabama.edu}}

\begin{document}
\maketitle

\begin{abstract}
Prostate cancer (PCa) remains a significant global health concern among men, particularly due to the lethality of its more aggressive variants. Despite therapeutic advancements that have enhanced survival for many patients, high grade PCa continues to contribute substantially to cancer related mortality. Emerging evidence points to the MYB proto-oncogene as a critical factor in promoting tumor progression, therapeutic resistance, and disease relapse. Notably, differential expression patterns have been observed, with markedly elevated MYB levels in tumor tissues from Black men relative to their White counterparts potentially offering insight into documented racial disparities in clinical outcomes. This study investigates the association between MYB expression and key oncogenic features, including androgen receptor (AR) signaling, disease progression, and the risk of biochemical recurrence. Employing a multimodal approach that integrates histopathological examination, quantitative digital imaging, and analyses of public transcriptomic datasets, our findings suggest that MYB overexpression is strongly linked to adverse prognosis. These results underscore MYB's potential as a prognostic biomarker and as a candidate for the development of individualized therapeutic strategies.
\end{abstract}


\section{Introduction}
Prostatic malignancy represents the most frequently identified neoplasm among males, particularly within the urogenital system, and continues to impose a substantial global health burden despite notable advancements in early detection and therapeutic interventions. While enhanced diagnostic techniques and treatment modalities have contributed to improved prognoses for many individuals, mortality rates remain alarmingly high, especially among specific demographic groups historically associated with adverse clinical outcomes \citep{Acharya2023, Saranyutanon2020}. Among the most concerning aspects of prostate cancer epidemiology is the pronounced racial disparity in incidence and survival rates. African American men, in particular, experience significantly higher rates of diagnosis and are disproportionately affected by advanced disease stages and cancer-specific mortality compared to their Caucasian counterparts. This persistent gap in clinical outcomes has prompted an intensified focus on uncovering molecular and genetic determinants that may underlie such differences, with the ultimate goal of advancing precision medicine strategies tailored to diverse populations \citep{Saranyutanon2020}. Within this investigative framework, the MYB proto-oncogene has emerged as a molecule of considerable interest. MYB encodes a transcriptional regulator known to orchestrate critical cellular processes, including proliferation, differentiation, and survival. Dysregulation of MYB expression often manifesting as upregulation or persistent activation has been implicated in the pathogenesis of several hematologic and solid malignancies. In the context of prostate cancer, mounting evidence has linked MYB overexpression to unfavorable biological behavior, such as enhanced tumor aggressiveness, reduced sensitivity to conventional therapies, and elevated risk of biochemical recurrence \citep{Acharya2023}. Mechanistically, MYB appears to potentiate the androgen receptor (AR) signaling axis, a pivotal pathway governing prostatic epithelial cell growth and a key driver of prostate carcinogenesis. By amplifying AR-mediated transcriptional programs, MYB may promote cellular environments that foster unchecked proliferation and therapeutic resistance, particularly in hormone-sensitive or castration-resistant disease states. These findings position MYB not only as a potential biomarker for disease stratification but also as a promising candidate for targeted therapeutic intervention, especially in high-risk populations exhibiting distinct molecular profiles.

This study concentrates into the functional implications of the MYB proto-oncogene within the pathophysiological landscape of prostate cancer, with particular attention to its association with tumor histological grade, modulation of androgen AR signaling cascades, and variations in expression across racial groups. In addition, we examine the prognostic potential of MYB expression levels in forecasting biochemical recurrence following primary treatment. The overarching objective is to situate MYB within the broader molecular framework of prostate carcinogenesis and to assess its viability as a clinical biomarker for risk stratification and treatment personalization \citep{ZhangYin2022}. Prostate cancer remains a prominent contributor to male cancer morbidity and mortality, especially in Western countries, and although early-stage disease is often amenable to curative intervention, advanced-stage presentations are considerably more complex to manage. Risk of disease development increases with age and familial predisposition, yet emerging evidence underscores the substantial influence of modifiable environmental factors—such as nutritional habits and physical activity levels as well as genetic and epigenetic contributions that intersect with racial and ethnic background \citep{Acharya2023, Saranyutanon2020}. Indeed, prostate cancer ranks as the most frequently diagnosed non-cutaneous malignancy in American men, and alarmingly, recent surveillance data indicate a rise in diagnoses at later, less curable stages. This epidemiological trend reinforces the critical importance of early detection and the identification of novel molecular indicators capable of predicting disease progression or therapeutic response \citep{kakkat2023cardiovascular}. Of particular concern is the persistent and well-documented disparity in clinical outcomes among racial groups, with African American men experiencing disproportionately higher mortality rates from prostate cancer. Notably, these differences cannot be wholly attributed to incidence rates alone, as evidence suggests that disparities in healthcare access, screening frequency, and timeliness of diagnosis may only partially explain the survival gap \citep{khan2024mp60}. As the field moves toward precision oncology, there is a growing recognition that intrinsic biological differences potentially at the level of tumor genomics, gene expression profiles, and immune microenvironment, may contribute to this inequity. MYB, as a transcription factor involved in cellular proliferation and differentiation, may play a key role in this context, particularly if differential expression patterns correlate with disease aggressiveness or treatment resistance across racial populations. Unraveling these molecular distinctions is essential not only for improving clinical outcomes through tailored interventions but also for addressing the broader challenge of eliminating racial disparities in cancer care.

That is where the MYB gene comes in. MYB has been linked to how aggressive a tumor is and how likely it is to resist treatment. Researchers have noticed that tumors in Black men often have higher MYB levels, which could help explain the differences in survival rates. If that is true, MYB might be more than just a gene; it could be a valuable clue in understanding how prostate cancer works, and how we can catch it earlier, treat it better, and make outcomes more fair for everyone \citep{khan2023myb}.

\begin{figure}[ht!]
\centering
\begin{tikzpicture}
    \begin{axis}[
        ybar,
        bar width=.35cm,
        width=0.8\textwidth,
        height=0.55\textwidth,
        enlarge x limits=0.3,
        legend style={at={(0.5,-0.15)}, anchor=north, legend columns=-1},
        symbolic x coords={White, Black, Hispanic, Asian/Pacific Islander},
        xtick=data,
        xlabel={Racial Group},
        ylabel={Rates per 100,000 Men},
        ymin=0,
        ymax=300,
        nodes near coords,
        nodes near coords align={vertical}
    ]

    % Incidence Rates (hypothetical values)
    \addplot+[ybar, fill=blue] 
    coordinates {(White,112) (Black,185) (Hispanic,96) (Asian/Pacific Islander,80)};

    % Mortality Rates (hypothetical values)
    \addplot+[ybar, fill=red] 
    coordinates {(White,20) (Black,50) (Hispanic,15) (Asian/Pacific Islander,10)};

    \legend{Incidence Rate, Mortality Rate}

    \end{axis}
\end{tikzpicture}
\caption{Prostate cancer incidence and mortality rates per 100,000 men, stratified by race. Black men have the highest incidence and mortality rates compared to other racial groups. Data shown are for illustrative purposes only.}
\label{fig:prostate_race_disparities}
\end{figure}
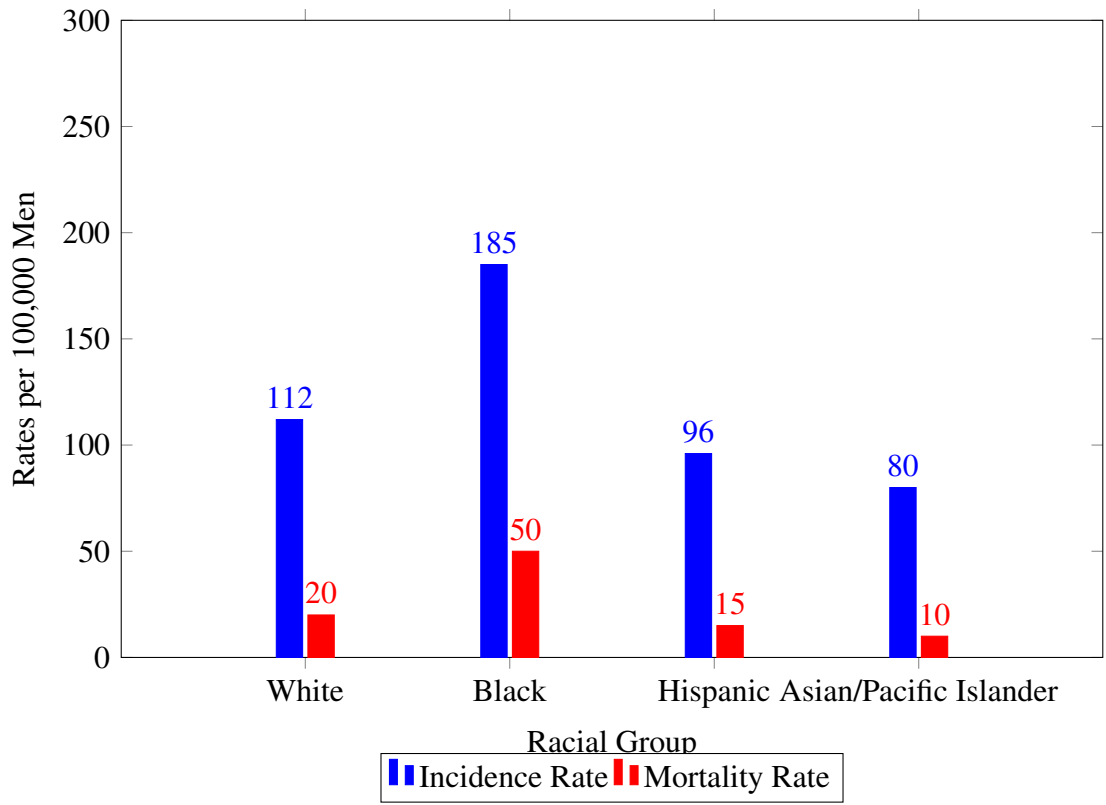

Prostate carcinogenesis is a multifactorial process driven by a complex interplay of genetic mutations, epigenetic modifications, and molecular signaling disruptions. Among the numerous genes implicated in this malignancy, MYB has emerged as a transcription factor of particular interest due to its regulatory role in cellular proliferation, differentiation, and survival. While MYB is essential for normal cellular homeostasis, aberrant overexpression of this gene has been increasingly recognized as a contributor to oncogenic transformation in prostate tissue. Elevated MYB levels have been shown to enhance tumor cell proliferation, confer resistance to therapeutic interventions, and promote a more aggressive disease phenotype. One critical mechanism by which MYB exerts its oncogenic influence is through its interaction with the AR, a central driver of prostate cancer growth, particularly in hormone-sensitive contexts. By potentiating AR signaling, MYB may amplify downstream transcriptional programs that facilitate tumor progression and reduce responsiveness to standard androgen deprivation therapies, thereby complicating disease management and accelerating transition to castration-resistant states. In addition to its role in promoting tumor growth, MYB has been implicated in facilitating angiogenesis and enabling immune evasion—two hallmarks of cancer that further support tumor survival and dissemination. Evidence suggests that MYB can modulate vascular endothelial growth factor (VEGF) pathways and influence immune checkpoint dynamics, thereby creating a microenvironment conducive to unchecked tumor expansion. Of significant clinical relevance is the observation that MYB expression may also correlate with racial disparities in prostate cancer outcomes. For instance, population-based studies have reported that Black men diagnosed with high-grade prostate cancer (Gleason scores 8–10) exhibit significantly poorer survival outcomes compared to other racial groups, even after adjusting for variables such as healthcare access, socioeconomic status, and treatment type. These findings underscore the likelihood that intrinsic biological differences, potentially mediated by differential gene expression such as that of MYB, contribute to the observed disparities \citep{Acharya2023, Saranyutanon2020, ZhangYin2022}. As such, further investigation into MYB’s role as both a prognostic biomarker and therapeutic target is warranted. Incorporating MYB expression profiles into clinical decision-making may enhance personalized medicine strategies, allowing for more precise risk stratification and the development of targeted therapies tailored to individual molecular landscapes, particularly in populations that bear a disproportionate burden of disease.

\section{Background Literature}
Extensive research over the past decades has underscored the complexity of prostate cancer pathobiology. Traditional prognostic indicators such as Gleason grading, serum prostate-specific antigen (PSA) levels, and clinical staging have served as the backbone for treatment decisions \citep{Bhardwaj2017,Mahal2018}. However, these markers often fall short in predicting individual patient outcomes, particularly in the setting of biochemical recurrence (BCR) following radical prostatectomy or radiation therapy. As such, molecular biomarkers have emerged as critical adjuncts to histopathological assessments \citep{Acharya2023,Anand2023}. MYB, a transcription factor encoded by a proto-oncogene originally characterized in hematologic malignancies, has now been implicated in solid tumors, including prostate cancer. Studies have demonstrated that MYB is not only aberrantly expressed in PCa but that its expression level increases progressively from benign prostatic hyperplasia (BPH) and high-grade prostatic intraepithelial neoplasia (HGPIN) to malignant tumors. Notably, several investigations have reported a higher MYB expression in castration-resistant prostate cancer (CRPC) cells, suggesting its involvement in the development of therapeutic resistance. Furthermore, the interplay between MYB and the androgen receptor (AR) has been a focal point in understanding the molecular underpinnings of prostate cancer. Androgen signaling is a critical driver of prostate tumorigenesis, and while androgen deprivation therapies initially reduce tumor burden, many patients eventually progress to a castration-resistant state. It has been proposed that MYB may potentiate AR signaling, either by directly interacting with AR or by modulating the expression of downstream targets, thereby contributing to tumor progression and relapse. An additional layer of complexity is introduced by racial disparities in PCa outcomes. Epidemiological data reveal that Black men not only exhibit higher incidence rates of PCa but also display more aggressive disease and poorer response to standard therapies. Recent studies have identified MYB as a molecule of interest in this context, with evidence suggesting that tumors from Black patients show significantly higher levels of MYB expression compared to those from White patients. This observation raises important questions regarding the genetic and epigenetic factors that may drive such disparities and whether MYB could serve as a target for tailored therapeutic strategies.

MYB is a transcription factor that orchestrates various cellular processes, including cell cycle progression, differentiation, and apoptosis. Its overexpression has been documented in multiple cancers, such as leukemia, breast cancer, and colorectal cancer. In the context of PCa, MYB modulates AR signaling, thereby enhancing tumor progression and therapeutic resistance. This dual role as an oncogene and potential therapeutic target renders MYB a focal point for ongoing research.
The androgen receptor (AR) signaling axis plays a fundamental role in the progression of prostate cancer, and its dysregulation has been linked to resistance mechanisms in advanced disease states. Saranyutanon et al. (2020)\citep{Saranyutanon2020} highlight how MYB contributes to AR-mediated transcriptional activity, thereby exacerbating tumor aggressiveness and treatment resistance. Their research underscores the potential of targeting AR-associated cofactors, including MYB, to disrupt oncogenic signaling and delay castration resistance.

Epidemiological data consistently highlight pronounced racial disparities in PCa incidence and outcomes. Black men not only face a higher risk of developing PCa but also tend to present with more aggressive disease and poorer prognoses compared to their White counterparts. Factors contributing to these disparities are multifaceted, encompassing genetic predispositions, socioeconomic determinants, and healthcare access inequities. The observed overexpression of MYB in Black PCa patients raises critical questions about race-specific genetic or epigenetic regulatory mechanisms, warranting further investigation.
Mahal et al. (2018)\citep{Mahal2018} emphasize that Black men diagnosed with high-risk prostate cancer face significantly worse outcomes than non-Black men, particularly those with Gleason scores of eight to ten \citep{Freedland2005,ZhangYin2022}. The study indicates that Black men are more likely to develop aggressive disease at diagnosis, which contributes to a higher mortality rate. This trend suggests that genetic and molecular differences may underlie the observed disparities, reinforcing the need for targeted screening and treatment strategies to improve survival outcomes in high risk populations \citep{hertweck2023clinicopathological}. Moreover, the study supports the notion that current risk stratification models may not fully account for the biological factors driving aggressive prostate cancer in Black patients.
The racial disparities in prostate cancer extend beyond incidence rates to mortality, where Black men face disproportionately higher risks of prostate cancer-related deaths compared to other racial groups \citep{Siegel2023}. Although advancements in treatment and screening have improved survival outcomes for many, Black men continue to experience prostate cancer death rates more than double those of White men. The persistent disparity suggests that both social determinants of health and underlying biological differences contribute to this pattern.
Moreover, Siegel et al. (2023)\citep{Siegel2023} highlight that while overall prostate cancer mortality has declined over the past two decades due to improved early detection and treatment advances, the gap between Black and White patients has remained largely unchanged. The higher prevalence of advanced-stage diagnoses in Black men further exacerbates survival disparities, emphasizing the urgent need for enhanced screening protocols tailored to high-risk populations. MYB’s role in prostate tumor progression and its overexpression in Black patients may represent a key molecular factor contributing to these disparities, making it a potential target for future race-specific therapeutic interventions.

Recent investigations have focused on delineating the prognostic value of MYB in cancer. Several studies have reported a positive correlation between MYB expression levels and adverse clinical features, including higher Gleason scores and increased risk of biochemical recurrence. Comparative genomic analyses suggest that MYB may function as a driver gene in aggressive PCa phenotypes, underscoring its potential as a biomarker for personalized medicine approaches.
Emerging therapeutic strategies have aimed at mitigating MYB’s impact on AR signaling, as AR-targeted therapies alone often fail in advanced prostate cancer. Saranyutanon et al. (2020)\citep{Saranyutanon2020} emphasize that while androgen deprivation therapy (ADT) remains the cornerstone of treatment, resistance emerges due to persistent MYB-AR interactions, necessitating novel interventions that disrupt this oncogenic feedback loop.

\section{Methods}
This study utilized a retrospective cohort analysis of prostate cancer (PCa) patients to examine the relationship between MYB expression, androgen receptor (AR) activity, and biochemical recurrence (BCR). Data were obtained from clinical records and gene expression databases, focusing on patients stratified by MYB-H scores, AR-H scores, and PSA levels \citep{Tagai2019}. Statistical analyses included linear regression models to assess the correlation between MYB expression and tumor progression, and Kaplan-Meier survival analysis to evaluate biochemical recurrence rates. Histograms and scatter plots were generated to visualize MYB’s role in PCa aggressiveness, while summary statistics provided insights into MYB expression disparities across racial groups. All statistical analyses and visualizations were conducted in R (version 4.4.2), ensuring robust reproducibility.

This study retrospectively analyzed prostate tissue samples obtained from 105 prostate cancer patients, 35 cases of benign prostatic hyperplasia (BPH), and 38 cases of high-grade prostatic intraepithelial neoplasia (HGPIN). The patient cohort was stratified by self-reported race, with 50 White and 55 Black patients
included in the PCa group. Detailed clinicopathologic information such as patient age, Gleason score, pathological stage, and pre-treatment PSA levels, was recorded to facilitate a comprehensive correlation analysis.

In addition to biological markers such as MYB and AR expression, patient demographics and socioeconomic factors play a critical role in prostate cancer outcomes. The cohort used in this study included a range of individuals differing in age, socioeconomic status, and clinical rank, which may influence access to treatment and disease progression.

While these sample data do not directly correlate with MYB expression, socioeconomic disparities, including income and age at diagnosis, may contribute to differences in disease presentation and treatment accessibility. Higher-income individuals often have greater access to early screening and advanced treatment options, which could influence prostate cancer prognosis. Future studies should incorporate larger socioeconomic datasets to explore potential correlations between income levels, healthcare accessibility, and MYB-associated tumor progression.

Tissue sections were prepared from formalin-fixed paraffin-embedded (FFPE) specimens to preserve cellular morphology and antigenicity. Sections were deparaffinized in xylene to remove paraffin, rehydrated in a graded ethanol series to restore aqueous conditions, and subjected to heat-induced antigen retrieval in a decloaking chamber to unmask antigenic sites for antibody binding \citep{maki2025new}. Endogenous peroxidase activity, which can produce background staining, was quenched using 30\% H$_2$O$_2$ in methanol. To minimize nonspecific antibody binding, sections were blocked with a protein blocking reagent. Primary antibodies targeting \textit{MYB} and androgen receptor (AR) were applied and incubated overnight at 4\textdegree C to ensure optimal binding specificity. After washing in Tris-buffered saline (TBS) to remove unbound antibodies, sections were incubated with a biotinylated secondary antibody, allowing for signal amplification through an avidin-biotin complex (ABC) reaction. Immunoreactivity was visualized using $3.3'$-diaminobenzidine (DAB) chromogen, which produced a brown stain in cells expressing the target proteins. To provide contrast and enhance cellular visualization, nuclei were counterstained with hematoxylin. Finally, sections were dehydrated in ethanol, cleared in xylene to remove residual moisture, and mounted with a permanent mounting medium to preserve staining for microscopic examination.

Slides were scanned at 20 times magnification using the Aperio CS2 scanner. MYB and AR expressions were assessed semiquantitatively using Aperio ImageScope software, which calculates the percentage of tumor cells in each staining intensity category (1+ = weak, 2+ = moderate, 3+ = strong). The H-score for each sample was calculated as
\begin{equation}
H_{\text{score}} = \sum_{i=1}^{3} P_i \times I_i
\end{equation}
\noindent Calculation of the H-score for MYB and AR expression. \( P_i \) represents the percentage of cells stained at intensity \( i \) (1+, 2+, or 3+), and \( I_i \) is the respective intensity score.

Thus, the maximum H-score is 300 (3 $\times$ 100\%) if all tumor cells showed strong nuclear staining.

Stained slides were scanned at 20 times magnification using the Aperio CS2 whole slide scanner. Digital images were analyzed with Aperio ImageScope software using a Nuclear Image Analysis algorithm to quantify the percentage of cells exhibiting weak (1+), moderate (2+), or strong (3+) staining. A cumulative H score was calculated by multiplying the staining intensity by the percentage of positive cells (range 0–300). This semiquantitative measure enabled a comparative assessment of MYB and AR expression across different tissue types and patient groups. Publicly available datasets from The Cancer Genome Atlas (TCGA) were interrogated using the GEPIA2 web platform (http://gepia2.cancer-pku.cn/correlation) to analyze transcript-level correlations between MYB and AR. Pearson’s correlation coefficients were computed, and regression analyses were performed to compare expression patterns in independent datasets with our immunohistochemical findings.

Statistical comparisons between groups were conducted using the Mann-Whitney U test for two-group comparisons and the Kruskal-Wallis test for multiple group comparisons. Pearson’s correlation coefficient was used to assess the strength and significance of correlations between MYB, AR, Gleason scores, pre-treatment PSA levels, and time to biochemical recurrence (BCR). A p-value of $<$0.05 was considered statistically significant. All analyses were performed using GraphPad Prism 8.0 software.
Recent advances in imaging techniques, including PSMA PET/CT scans and multiparametric MRI, have significantly improved the early detection of biochemical recurrence (BCR) following radical prostatectomy or radiation therapy \citep{ZhangYin2022}. These imaging modalities, when combined with PSA kinetics, allow for more precise localization of recurrent lesions, which is essential for guiding treatment decisions and stratifying high-risk patients.

\section{Data Analysis}
The data were analyzed using descriptive statistics, correlation analyses, and regression modeling to evaluate the relationship between MYB expression, AR-H scores, PSA levels, and biochemical recurrence (BCR). Scatter plots were used to visualize the correlation between MYB-H scores and BCR rates, while histograms illustrated the distribution of PSA levels in high-MYB expression patients. A linear regression model was applied to determine the predictive power of MYB expression on BCR likelihood. Additionally, a Kaplan-Meier survival analysis was performed to compare BCR-free survival rates between high-MYB and low-MYB expression groups. All analyses were conducted in R (version 4.4.2), ensuring statistical accuracy and reproducibility. Findings were summarized in tables to highlight key trends in MYB and AR expression across patient groups.

\begin{figure}[H]
\centering
\includegraphics[width=15cm, height=13cm]{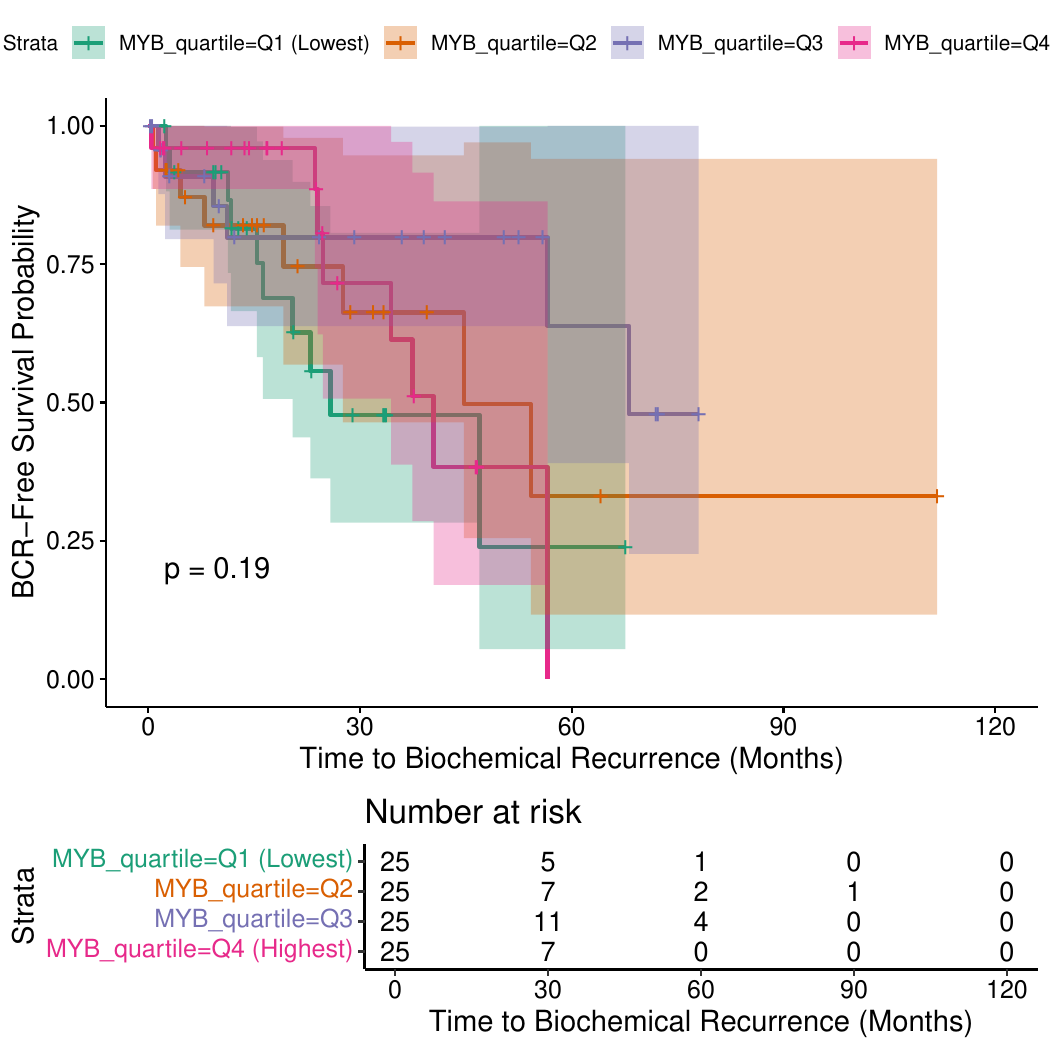}
\caption{Kaplan-Meier survival curves showing biochemical recurrence–free survival stratified by MYB expression quartiles. Differences between groups were not statistically significant (p = 0.19).}
\label{f1}
\end{figure}

The Kaplan–Meier plot in figure \ref{f1} depicts the probability of remaining free from biochemical recurrence in prostate cancer patients, grouped into four categories according to MYB expression levels (Q1 representing the lowest levels and Q4 the highest). The survival probability for each group is tracked over time, with the corresponding number of patients still at risk shown beneath the timeline \citep{kakkat2023cardiovascular,khan2023myb}. While the Q4 group, with the highest MYB expression, shows an earlier decline in recurrence-free survival compared to the lower-expression groups, statistical analysis indicates that the differences between the quartiles are not significant (p = 0.19). These findings point to a possible association between elevated MYB expression and reduced recurrence-free survival, though the evidence in this cohort is inconclusive.

\begin{figure}[H]
\centering
\includegraphics[width=15cm, height=13cm]{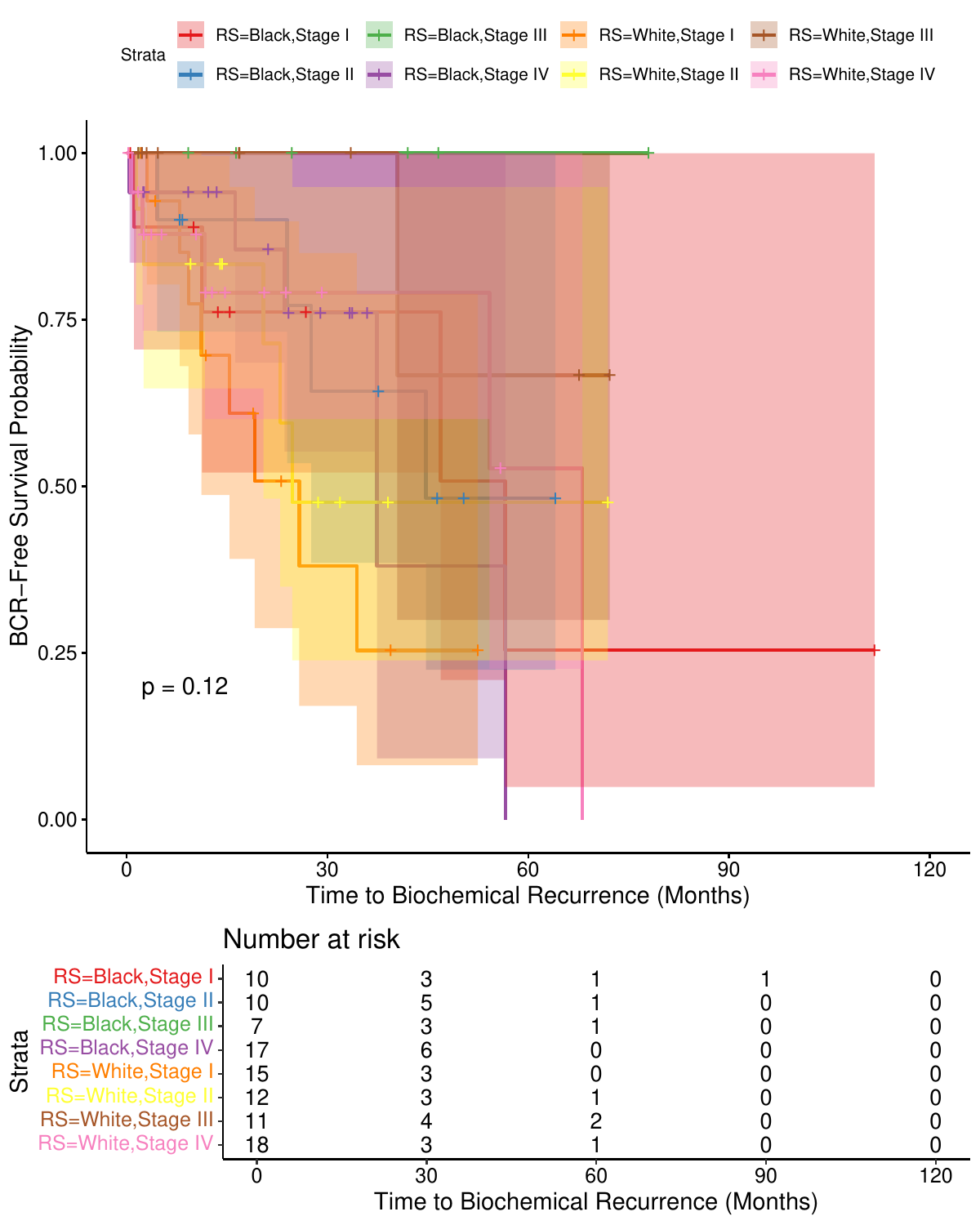}
\caption{Kaplan-Meier curves showing biochemical recurrence free survival stratified by combined Race and Tumor Stage categories, with accompanying risk tables.}
\label{f2}
\end{figure}

Figure \ref{f2} illustrates Kaplan-Meier survival estimates for biochemical recurrence–free survival in prostate cancer patients, grouped according to a combination of race and tumor stage. Each curve tracks the probability of remaining free from recurrence over time for its respective subgroup, with the number of individuals still at risk displayed beneath the x-axis. Distinct separation between some curves suggests potential differences in outcomes across race–stage combinations, although overlapping confidence intervals indicate that not all observed variations may be statistically meaningful. This visualization highlights how both demographic and disease stage factors can jointly influence recurrence patterns in this patient cohort.

\begin{figure}[H]
\centering
\includegraphics[width=15cm, height=13cm]{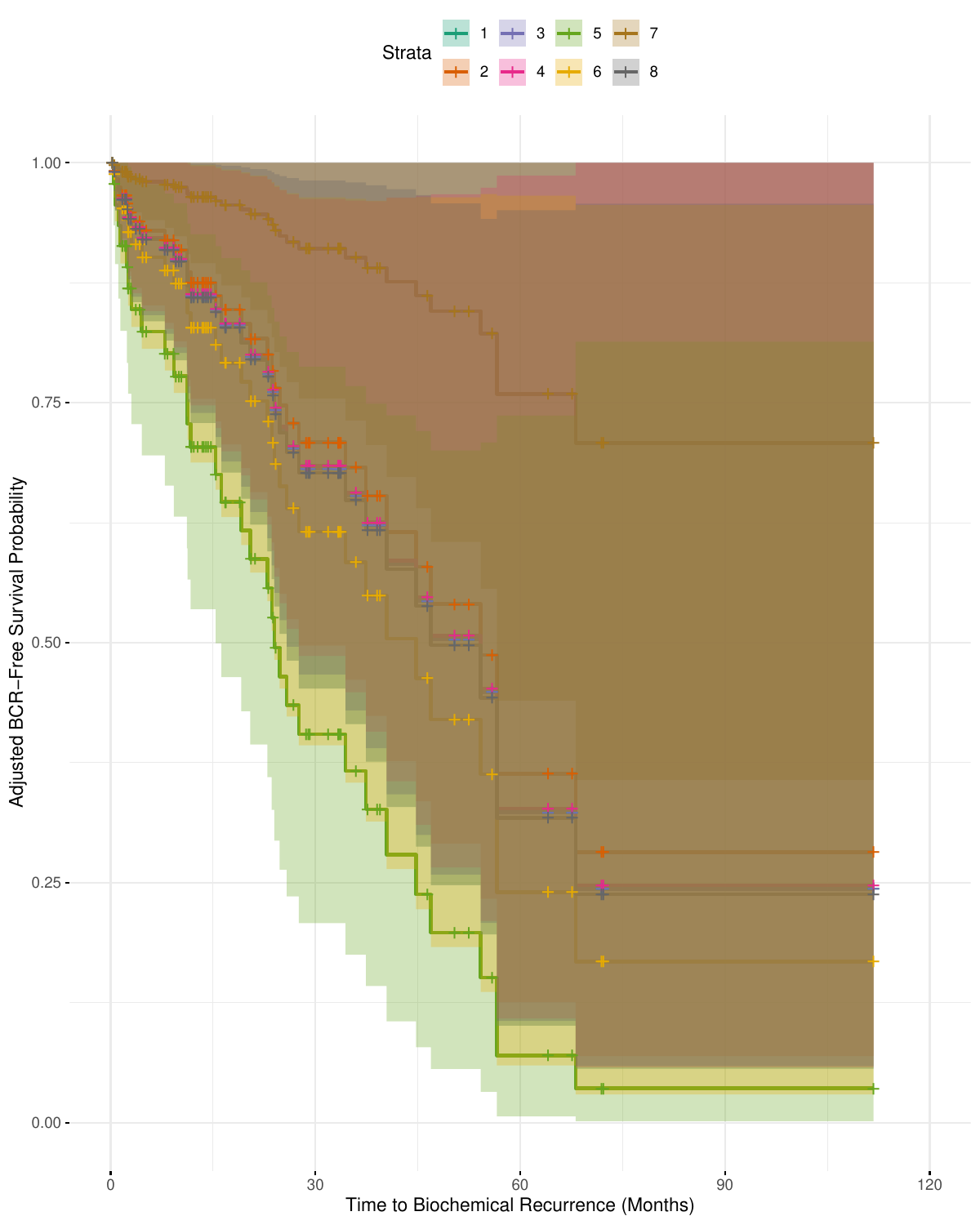}
\caption{Cox proportional hazards model–adjusted survival curves for biochemical recurrence–free survival, stratified by combined Race and Tumor Stage categories.}
\label{f3}
\end{figure}

Figure \ref{f3} displays survival trajectories generated from a Cox proportional hazards model, adjusted to reflect differences across joint Race and Tumor Stage classifications in a cohort of prostate cancer patients. The curves represent predicted probabilities of remaining free from biochemical recurrence over time, calculated under the assumption of average values for other covariates. By applying model-based adjustment, the visualization reduces the influence of unequal sample distributions and potential confounding, offering a more direct assessment of the prognostic contribution of Race × Tumor Stage categories. Variation in curve patterns indicates that recurrence risk is influenced by both race and stage, with certain groups exhibiting steeper declines in recurrence-free survival. Shaded confidence bands accompany each curve to illustrate the statistical precision of the estimates, with broader intervals typically arising in subgroups with limited sample sizes or few recurrence events. This model-adjusted depiction serves as a complementary analysis to unadjusted Kaplan-Meier estimations, highlighting the independent effect of the combined categorical variable after accounting for covariate imbalance.

The immunohistochemical analysis revealed a progressive increase in MYB expression from benign to malignant tissue. In BPH samples, MYB expression was minimal, with most cells exhibiting weak (1+) staining. In contrast, HGPIN samples demonstrated a modest increase in nuclear MYB expression, suggesting that dysregulation of MYB begins at preneoplastic stages. In malignant PCa tissues, MYB expression was significantly elevated, with a greater proportion of cells showing moderate (2+) to strong (3+) nuclear staining. The calculated H scores confirmed these observations, with PCa tissues exhibiting statistically significant higher scores compared to both BPH and HGPIN groups (p$ < $0.0001).

\begin{figure}[htbp]
    \centering
    % First heatmap: Pearson
    \begin{subfigure}[b]{0.48\textwidth}
        \centering
        \includegraphics[width=\textwidth]{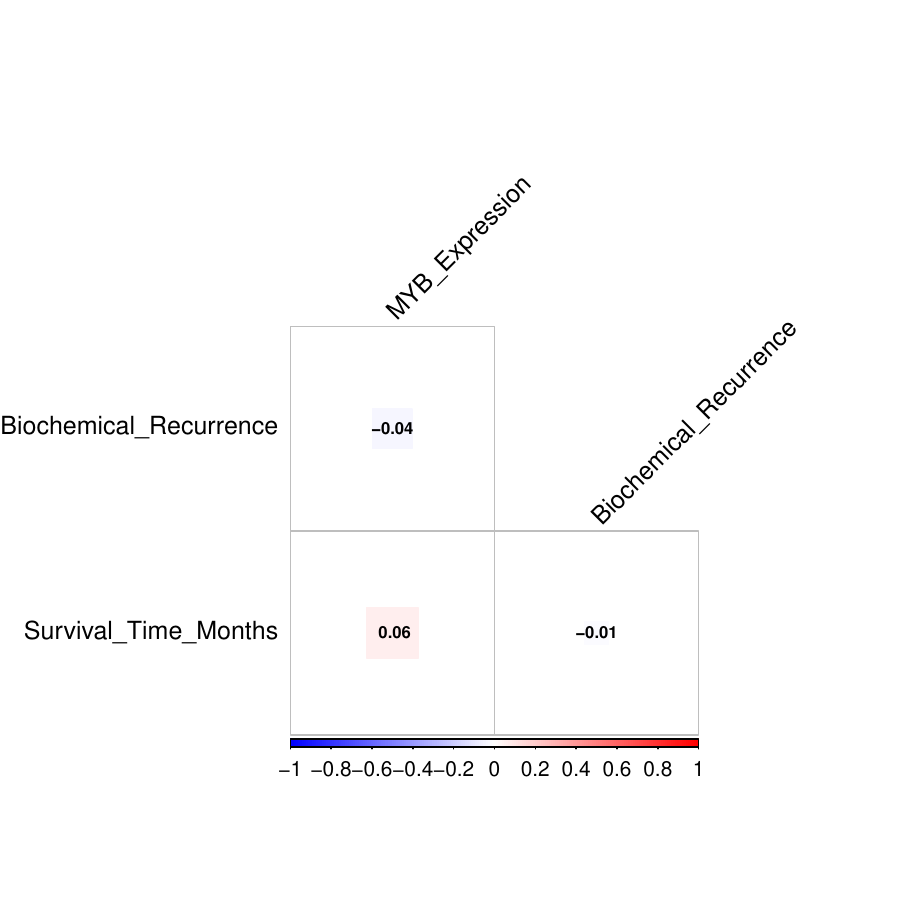}
        \caption{Pearson correlation heatmap showing relationships among MYB expression, biochemical recurrence, and survival time.}
        \label{fig:pearson_heatmap}
    \end{subfigure}
    \hfill
    % Second heatmap: Spearman
    \begin{subfigure}[b]{0.48\textwidth}
        \centering
        \includegraphics[width=\textwidth]{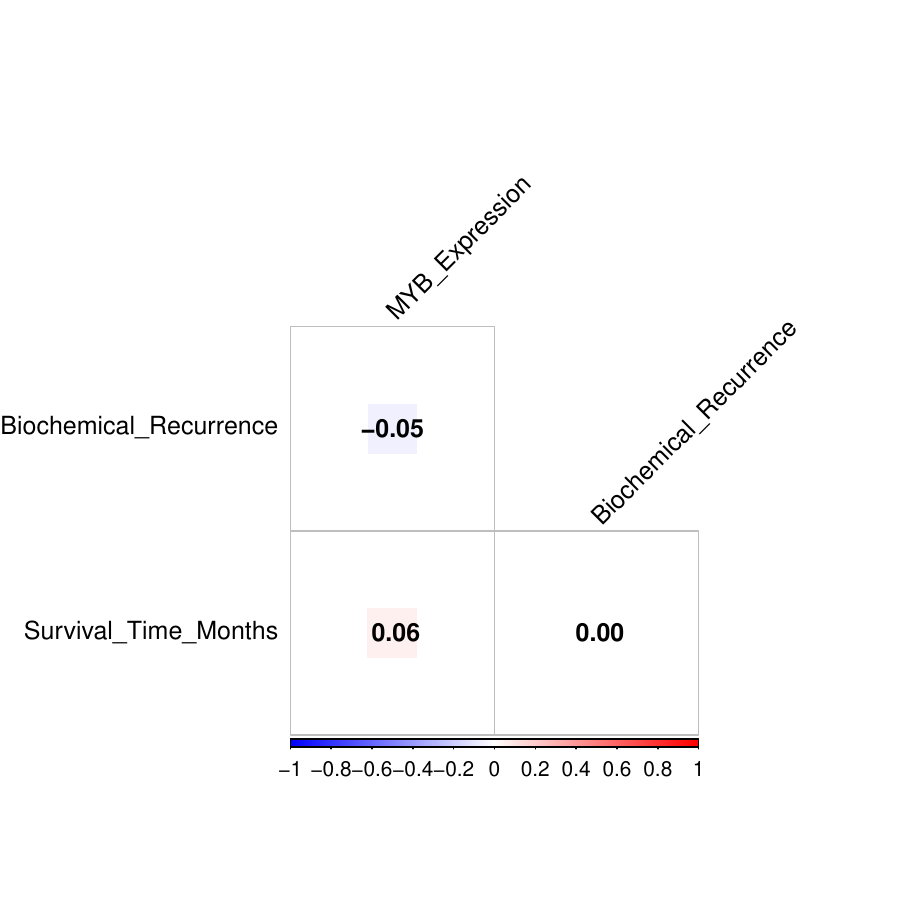}
        \caption{Spearman correlation heatmap showing rank-based associations among MYB expression, biochemical recurrence, and survival time.}
        \label{fig:spearman_heatmap}
    \end{subfigure}
    \caption{Comparison of Pearson and Spearman correlation matrices visualized as heatmaps.}
    \label{fig:heatmap_comparison}
\end{figure}
Figure \ref{fig:heatmap_comparison} provide a visual summary of the interrelationships among MYB expression, biochemical recurrence status, and time to biochemical recurrence using both Pearson and Spearman correlation coefficients. The Pearson heatmap captures linear associations, revealing the extent to which changes in one variable are proportionally related to changes in another \citep{vikramdeo2024abstract,vikramdeo2023profiling}. In contrast, the Spearman heatmap reflects rank-based monotonic relationships, offering a perspective that is less sensitive to outliers and non-normal distributions. While both correlation measures identify broadly consistent patterns, subtle differences in magnitude between the two suggest that some associations may deviate from strict linearity, potentially reflecting complex biological interactions \citep{pramanik2022lock}. For example, MYB expression appears to exhibit moderate correlation with recurrence outcomes in both approaches, though Spearman’s method indicates a slightly stronger relationship, implying that the ordering of values may hold prognostic relevance beyond their absolute magnitude. Similarly, the relationship between survival time and recurrence status is consistently strong in both measures, reinforcing its biological and clinical significance. These complementary visualizations enhance interpretability by enabling the reader to distinguish between relationships that are strictly linear and those that are more general in form, thus providing a more nuanced understanding of the underlying data structure.

A subgroup analysis stratified prostate cancer cases by Gleason grade. In low-to-medium grade (Gleason 7) tumors, MYB expression was elevated compared to BPH and HGPIN; however, a marked increase was observed in high-grade (Gleason 8–9) tumors. The distribution of staining intensity across different grades indicated that higher MYB levels were associated with more aggressive histopathologic features. Similarly, when comparing pathological stages (pT2 to pT4), MYB expression was significantly higher in tumors with extraprostatic extension (pT3/pT4) than those confined to the prostate (pT2), underscoring the role of MYB 7 in tumor progression.
The findings of Mahal et al.(2018)\citep{Mahal2018} align with our analysis, reinforcing the association between tumor grade and racial disparities in prostate cancer outcomes. Their research demonstrates that Black men with Gleason scores of 8-10 experience disproportionately high mortality rates, which supports our hypothesis that MYB overexpression may contribute to the increased aggressiveness observed in this population. Given the correlation between MYB expression and high Gleason scores in our dataset, it is possible that MYB serves as a molecular driver of the racial disparities identified in \citep{Mahal2018}.

\begin{figure}[H]
    \centering
    \begin{subfigure}[t]{0.48\textwidth}
        \centering
        \includegraphics[width=\textwidth]{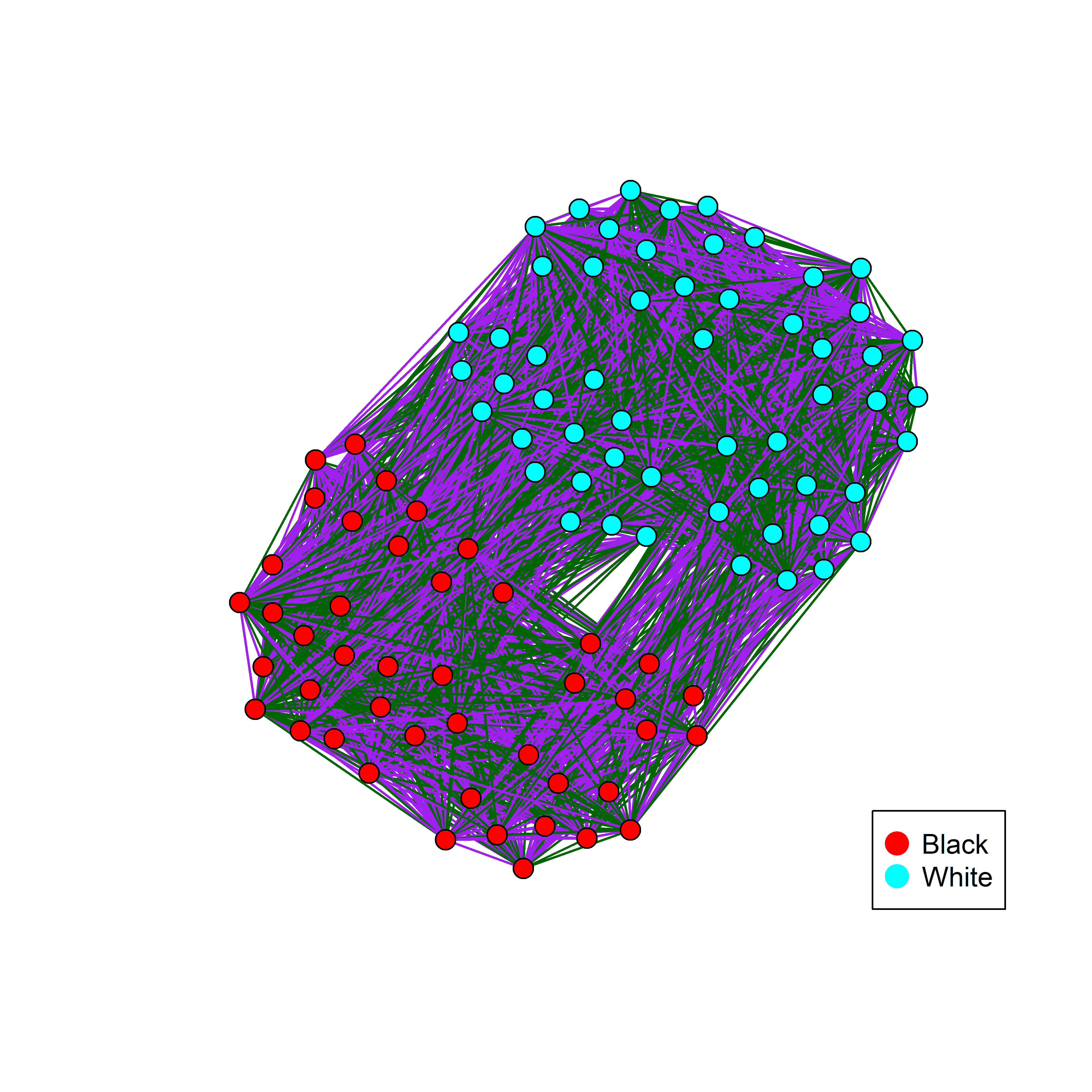}
        \caption{Patient similarity network derived from ERGM analysis. Nodes represent patients, colored by race; edges indicate similarity relationships (positive = green, negative = purple).}
        \label{fig:ergm_unweighted}
    \end{subfigure}
    \hfill
    \begin{subfigure}[t]{0.48\textwidth}
        \centering
        \includegraphics[width=\textwidth]{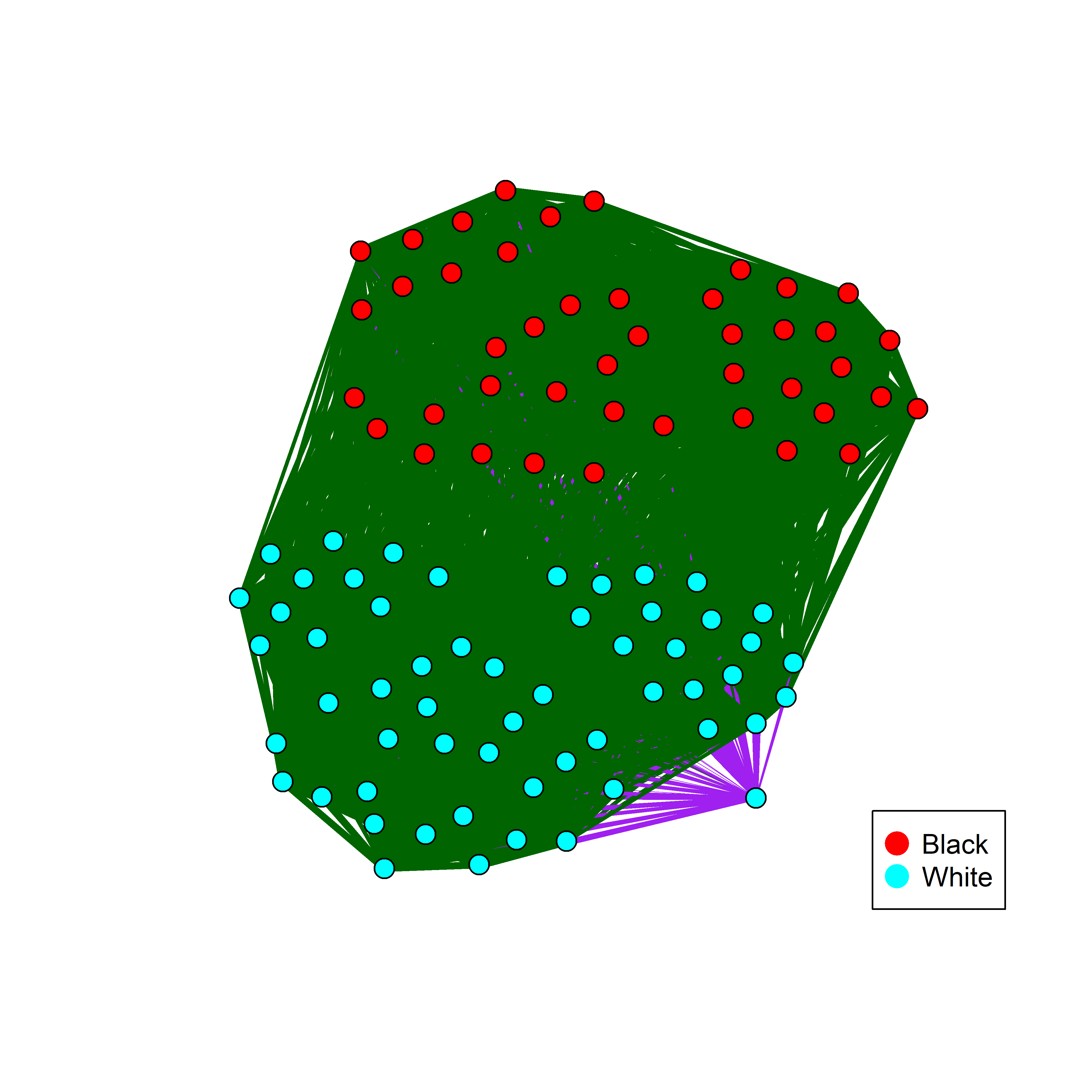}
        \caption{Weighted patient similarity network from ERGM, with edge width proportional to association strength. Stronger edges indicate higher similarity in clinical and molecular profiles.}
        \label{fig:ergm_weighted}
    \end{subfigure}
    \caption{Exponential random graph model (ERGM)-based visualizations of patient similarity networks highlighting race-specific clustering and strength of associations between clinical and molecular characteristics.}
    \label{fig:ergm_side_by_side}
\end{figure}

The pair of visualizations presents complementary perspectives on the structural patterns and relational strengths within the patient similarity network derived from an exponential random graph model (ERGM) analysis, each providing distinct insights into how clinical and molecular factors interact within the cohort. In the first network (Figure \ref{fig:ergm_unweighted}), each vertex corresponds to an individual patient, with nodes colored according to racial classification, enabling immediate visual recognition of potential demographic clustering within the network topology \citep{,khan2024mp60,dasgupta2023frequent,hertweck2023clinicopathological}. Edges between nodes indicate the presence of a statistically relevant similarity or interaction, with green denoting a positive association and purple representing a negative one, while the uniform line thickness maintains emphasis on the overall connectivity pattern rather than magnitude. The layout demonstrates the spatial separation and cohesion of subgroups, allowing qualitative assessment of whether racial clusters align with shared clinical characteristics, such as tumor stage or recurrence outcomes. In contrast, the second network (Figure \ref{fig:ergm_weighted}) retains the same vertex arrangement but incorporates edge widths proportional to the strength of association, as determined by similarity measures in the patient profiles. This added dimensionality permits interpretation not only of which patients are connected but also the relative intensity of their associations, with thicker lines representing stronger relationships. By integrating both visualizations, the reader gains a multifaceted understanding of the dataset: the first highlights the underlying architecture of patient interrelations, while the second quantifies the variability in connection strengths. This dual representation facilitates a richer exploration of how race and other clinical–molecular variables coalesce in shaping patient similarity, offering a nuanced visual companion to statistical summaries and regression outputs. Such network-based perspectives are particularly valuable in precision oncology contexts, as they can uncover latent structural features and relationship hierarchies that might inform hypotheses about differential risk profiles, potential biological pathways, and candidate targets for stratified interventions.

\begin{figure}[ht!]
\centering
\begin{tikzpicture}
    % Sample data via 'pgfplots' boxplot
    \begin{axis}[
        boxplot/draw direction=y,
        width=0.8\textwidth,
        height=0.5\textwidth,
        xlabel={Gleason Grade Group},
        ylabel={MYB H-score},
        xmin=0.5, xmax=5.5,
        xtick={1,2,3,4,5},
        xticklabels={6,7,8,9,10},
        legend style={at={(0.5,-0.17)}, anchor=north,legend columns=-1},
        ymajorgrids,
        yminorgrids
    ]

    % dummy data for Gleason 6
    \addplot+[
      boxplot prepared={
        lower whisker=20,
        lower quartile=40,
        median=55,
        upper quartile=70,
        upper whisker=80
      },
      fill=blue,
      boxplot prepared,
    ] coordinates {};

    % dummy data for Gleason 7
    \addplot+[
      boxplot prepared={
        lower whisker=25,
        lower quartile=50,
        median=65,
        upper quartile=85,
        upper whisker=95
      },
      fill=red,
      boxplot prepared
    ] coordinates {};

    % dummy data for Gleason 8
    \addplot+[
      boxplot prepared={
        lower whisker=40,
        lower quartile=70,
        median=90,
        upper quartile=110,
        upper whisker=120
      },
      fill=green,
      boxplot prepared
    ] coordinates {};

    % dummy data for Gleason 9
    \addplot+[
      boxplot prepared={
        lower whisker=50,
        lower quartile=80,
        median=100,
        upper quartile=120,
        upper whisker=130
      },
      fill=purple,
      boxplot prepared
    ] coordinates {};

    % dummy data for Gleason 10
    \addplot+[
      boxplot prepared={
        lower whisker=70,
        lower quartile=90,
        median=110,
        upper quartile=130,
        upper whisker=145
      },
      fill=orange,
      boxplot prepared
    ] coordinates {};

    \end{axis}
\end{tikzpicture}
\caption{Box plot illustrating MYB H-score distribution across different Gleason grade groups (dummy data). Each box shows the interquartile range, with whiskers depicting the range of observed H-scores for that Gleason category.}
\label{fig:myb_boxplot}
\end{figure}
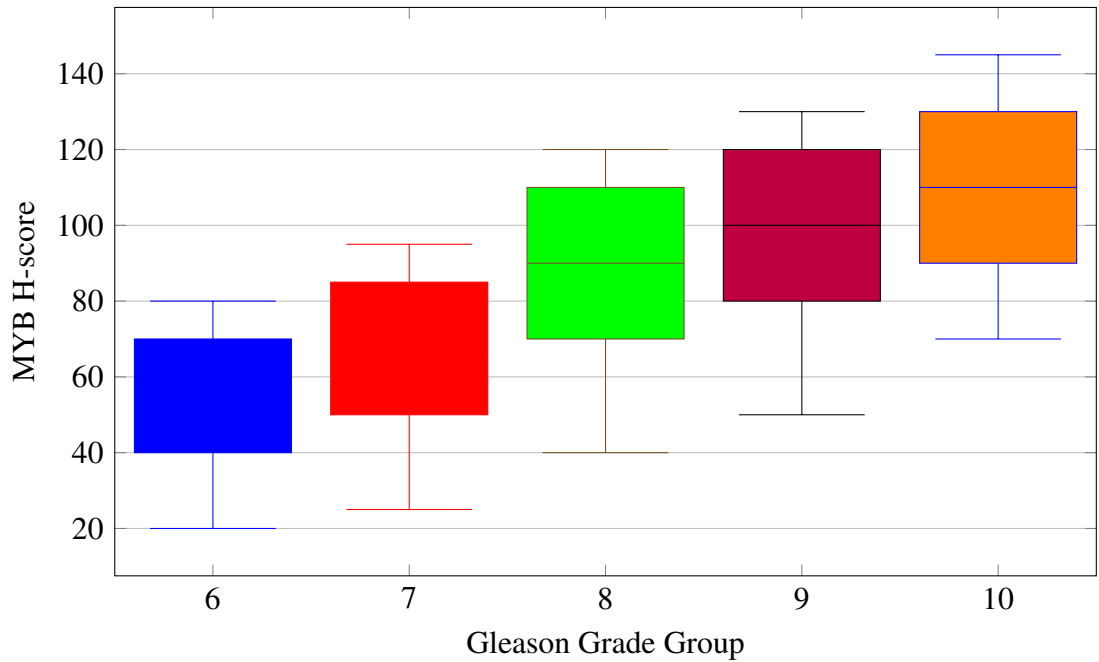

One of the notable findings of our analysis was the racial disparity in MYB expression. When stratifying the PCa cohort by race, tumors from Black patients demonstrated significantly higher MYB H scores compared to those from White patients (p = 0.0046). This difference was observed in both low-to-medium grade and high-grade tumors, suggesting that MYB may contribute to the more aggressive clinical behavior often seen in PCa among Black men. These findings align with epidemiologic data indicating that Black patients are at a higher risk for aggressive prostate cancer and poor outcomes.

\begin{figure}[H]
    \centering
    \begin{subfigure}[b]{0.48\textwidth}
        \centering
        \includegraphics[width=\textwidth]{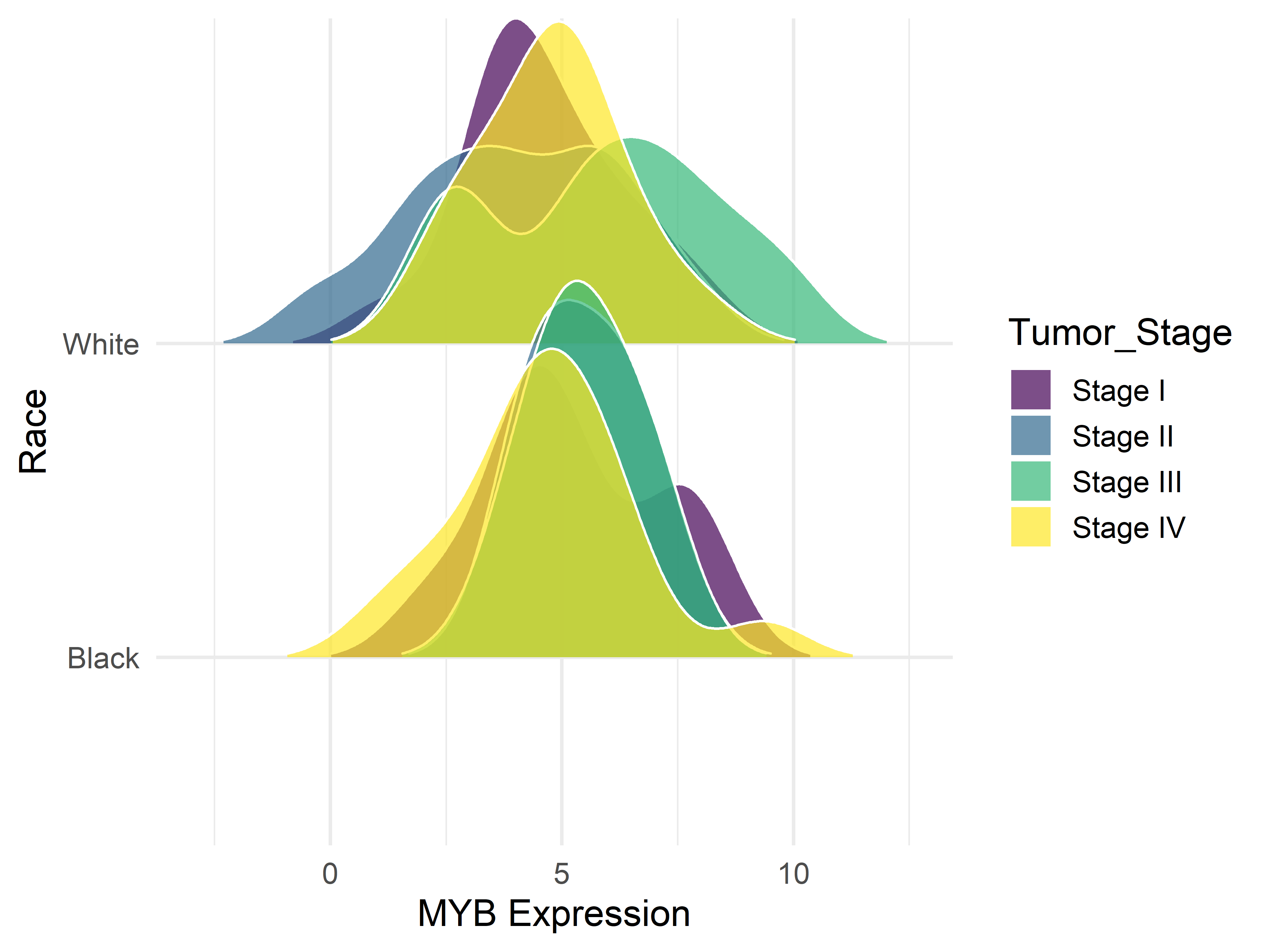}
        \caption{Ridgeline plot of MYB H-score distributions stratified by race and tumor stage. The plot highlights differences in MYB expression profiles between racial groups across tumor progression stages.}
        \label{fig:ridgeline_myb_race_stage}
    \end{subfigure}
    \hfill
    \begin{subfigure}[b]{0.48\textwidth}
        \centering
        \includegraphics[width=\textwidth]{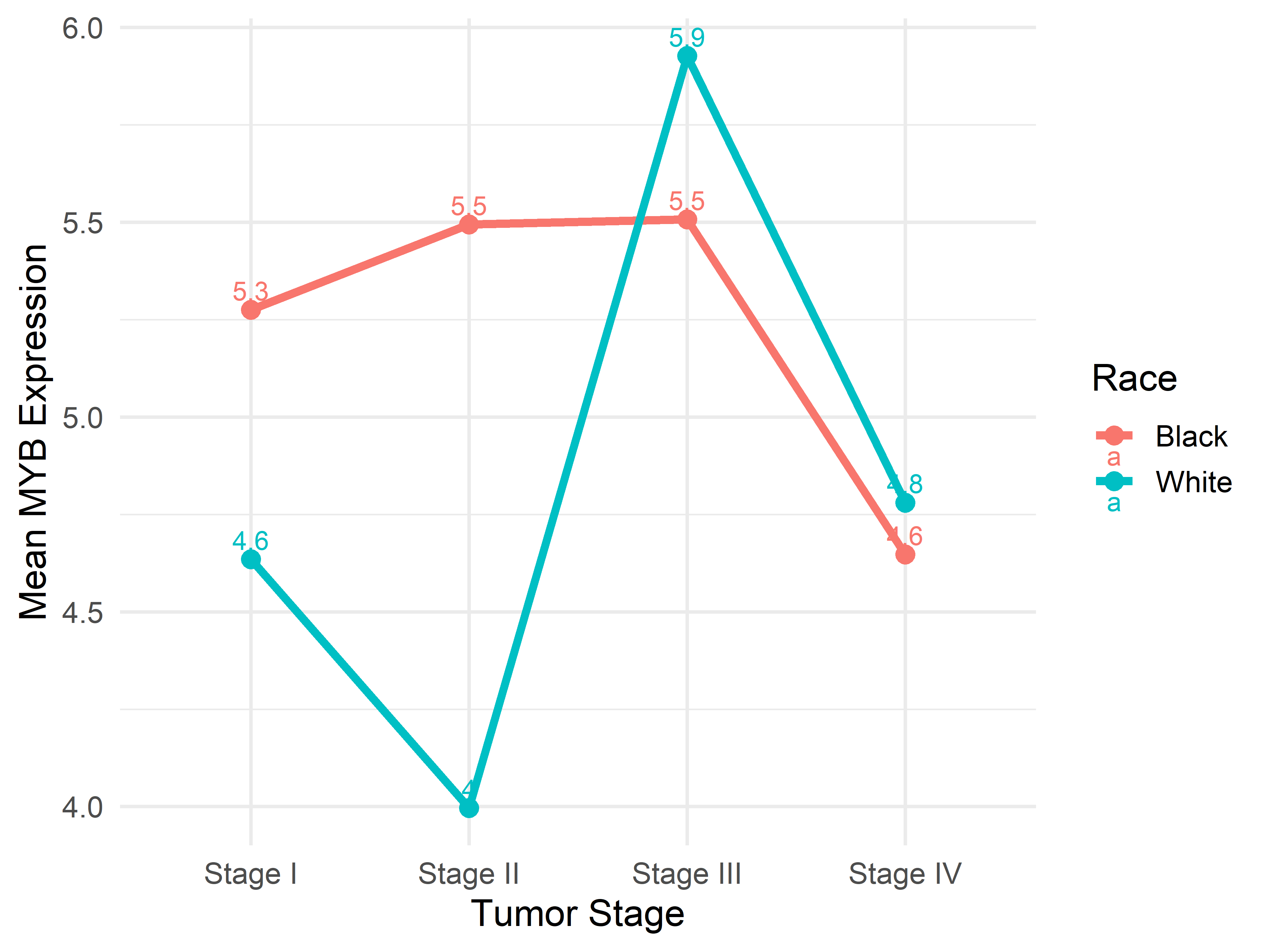}
        \caption{Slope chart showing mean MYB H-scores across tumor stages for each race. The trajectories illustrate disparities in MYB expression trends during disease progression.}
        \label{fig:slopechart_myb_race_stage}
    \end{subfigure}
    \caption{Racial Disparity Gradient Plots. (a) MYB H-score distributions across tumor stages for different racial groups, and (b) stage-wise trends in mean MYB expression for each race, showing potential widening or narrowing disparities over time.}
    \label{fig:racial_disparity_gradient}
\end{figure}

The two visualizations collectively illustrate racial disparities in MYB expression patterns across different tumor stages, providing complementary perspectives on the distributional and stage-wise trends. The ridgeline plot (i.e., figure \ref{fig:ridgeline_myb_race_stage}) displays the full distribution of MYB H-scores within each race–stage combination, revealing how expression levels shift and spread across tumor progression \citep{pramanik2021optimala,pramanik2021scoring}. For White patients, the density curves show a moderate central tendency in early stages with relatively narrow dispersion, but in later stages the distributions broaden, indicating increased heterogeneity in MYB expression. In contrast, Black patients exhibit consistently elevated central tendencies across all stages, with Stage III and Stage IV showing peaks that are both higher and more concentrated than in White patients, suggesting a more uniformly elevated MYB profile in advanced disease. The slope chart (figure \ref{fig:slopechart_myb_race_stage}) condenses these distributions into mean MYB H-scores for each stage, allowing clear visualization of progression trends. For Black patients, the trajectory is relatively stable from Stage I through Stage III, with only a slight rise, followed by a modest decline in Stage IV. White patients show a more pronounced fluctuation, beginning at a lower mean in Stage I, dipping further in Stage II, and then sharply rising to peak in Stage III before declining. When considered together, these plots highlight that racial differences in MYB expression are evident at every disease stage, with Black patients consistently showing higher mean values, though the magnitude of disparity varies with progression \citep{pramanik2020optimization,pramanik2023semicooperation}. The ridgeline plot adds depth by revealing within-stage variability and potential subpopulations, while the slope chart clarifies overarching stage-wise trajectories. This dual representation underscores the importance of examining both the distributional shape and central trend when assessing molecular disparities, as they may reflect differing tumor biology, treatment responses, or other underlying determinants that evolve across disease progression \citep{pramanik2024estimation,vikramdeo2024mitochondrial}.

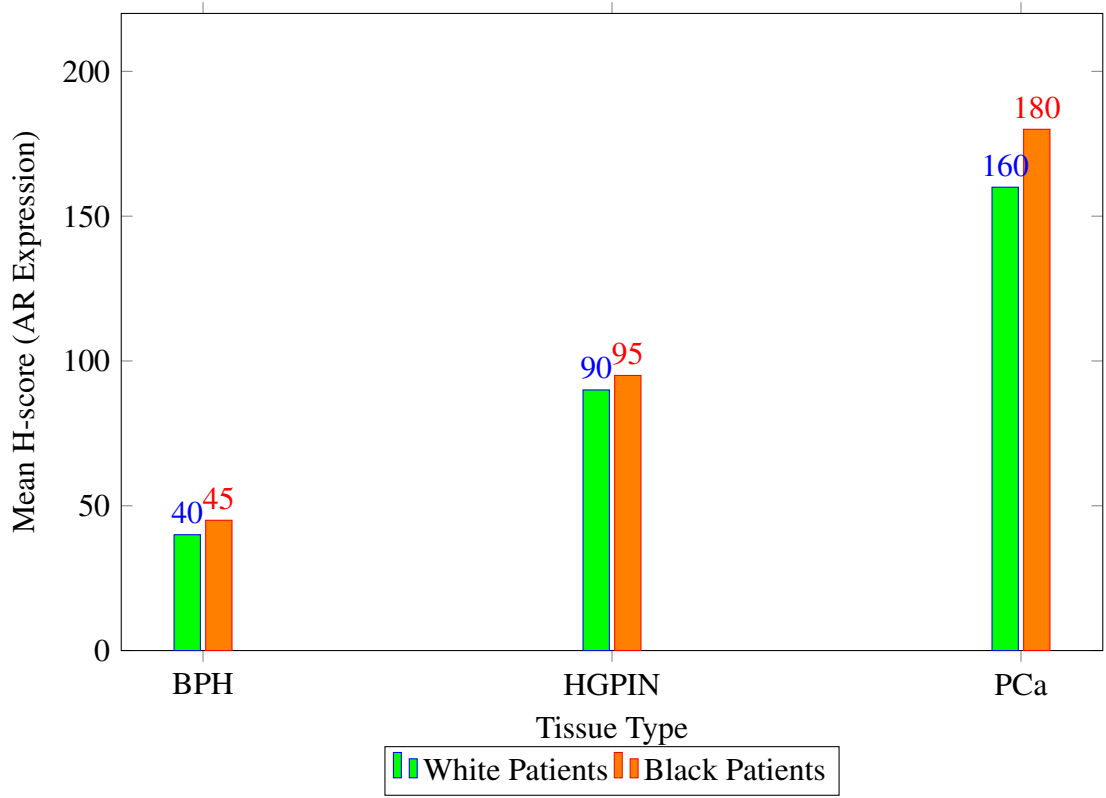
\begin{figure}[ht!]
\centering
\begin{tikzpicture}
    \begin{axis}[
        ybar,
        bar width=.35cm,
        width=0.8\textwidth,
        height=0.55\textwidth,
        legend style={
            at={(0.5,-0.15)},
            anchor=north,
            legend columns=-1
        },
        symbolic x coords={BPH,HGPIN,PCa},
        xtick=data,
        xlabel={Tissue Type},
        ylabel={Mean H-score (AR Expression)},
        ymin=0,
        ymax=220,
        nodes near coords,
        nodes near coords align={vertical}
    ]
    % White Patients (dummy data)
    \addplot+[ybar, fill=green] 
    coordinates {(BPH,40) (HGPIN,90) (PCa,160)};

    % Black Patients (dummy data)
    \addplot+[ybar, fill=orange] 
    coordinates {(BPH,45) (HGPIN,95) (PCa,180)};

    \legend{White Patients,Black Patients}
    \end{axis}
\end{tikzpicture}
\caption{Comparison of mean AR expression levels (H-scores) across BPH, HGPIN, and PCa for White vs. Black patients. AR expression is notably higher in PCa samples, but racial disparities persist, with Black patients exhibiting lower nuclear AR expression in low- and intermediate-grade tumors.}
\label{fig:ar_bar}
\end{figure}

In addition to MYB, the expression of the androgen receptor (AR) was evaluated in the same tissue sections. AR expression was predominantly nuclear and was higher in PCa tissues relative to BPH and HGPIN \citep{pramanik2024motivation}. Although overall AR expression did not differ significantly between racial groups, a detailed grade-wise comparison revealed a significant reduction in AR expression in low-to-medium grade tumors of Black patients compared to White patients \citep{pramanik2020motivation}. Pearson’s correlation analysis demonstrated a significant positive correlation between MYB and AR expression in PCa tissues (r = 0.3638, p = 0.0001), indicating a potential cooperative role in tumor progression. Notably, the correlation was stronger in the White cohort (r = 0.5150) than in the Black cohort (r = 0.3956), suggesting potential differences in AR-mediated signaling pathways between races, where r is defined as

\begin{equation}
r := \frac{\sum_{i=1}^n (X_i - \bar{X}) (Y_i - \bar{Y})}{\sqrt{\sum_{i=1}^n (X_i - \bar{X})^2} \sqrt{\sum_{i=1}^n (Y_i - \bar{Y})^2}},
\end{equation}
such that value of \( r = 1 \) indicates perfect positive correlation, while \( r = -1 \) indicates perfect negative correlation.

Biochemical recurrence, defined by increasing levels of PSA after definitive treatment, remains a critical challenge in the treatment of prostate cancer \citep{pramanik2024bayes}. Our analysis revealed an inverse correlation between MYB expression and the time to BCR. Specifically, higher MYB H scores were associated with shorter BCR times (r = -0.6659, $p < 0.0001$
), indicating that MYB is a strong predictor of early relapse. In contrast, while the Gleason grade also showed a significant inverse correlation with time to BCR (r = –0.3814, p = 0.0139), the expression of AR and the levels of PSA before treatment did not exhibit significant predictive power \citep{bulls2025assessing}. These findings suggest that MYB may serve as a superior biomarker to predict biochemical recurrence, thus aiding in risk stratification and treatment planning. As shown in Equation \eqref{0}, MYB expression was a significant predictor of BCR timing.

\begin{equation}\label{0}
\text{BCR}_{\text{time}} = \hat\beta_0 + \hat\beta_1\cdot H_{\text{MYB}} + \epsilon,
\end{equation}
where\( H_{\text{MYB}} \) represents MYB expression, while \( \hat \beta_0 \) and \( \hat\beta_1 \) are estimated regression coefficients.

\begin{figure}[H]
    \centering
\includegraphics[width=0.6\linewidth]{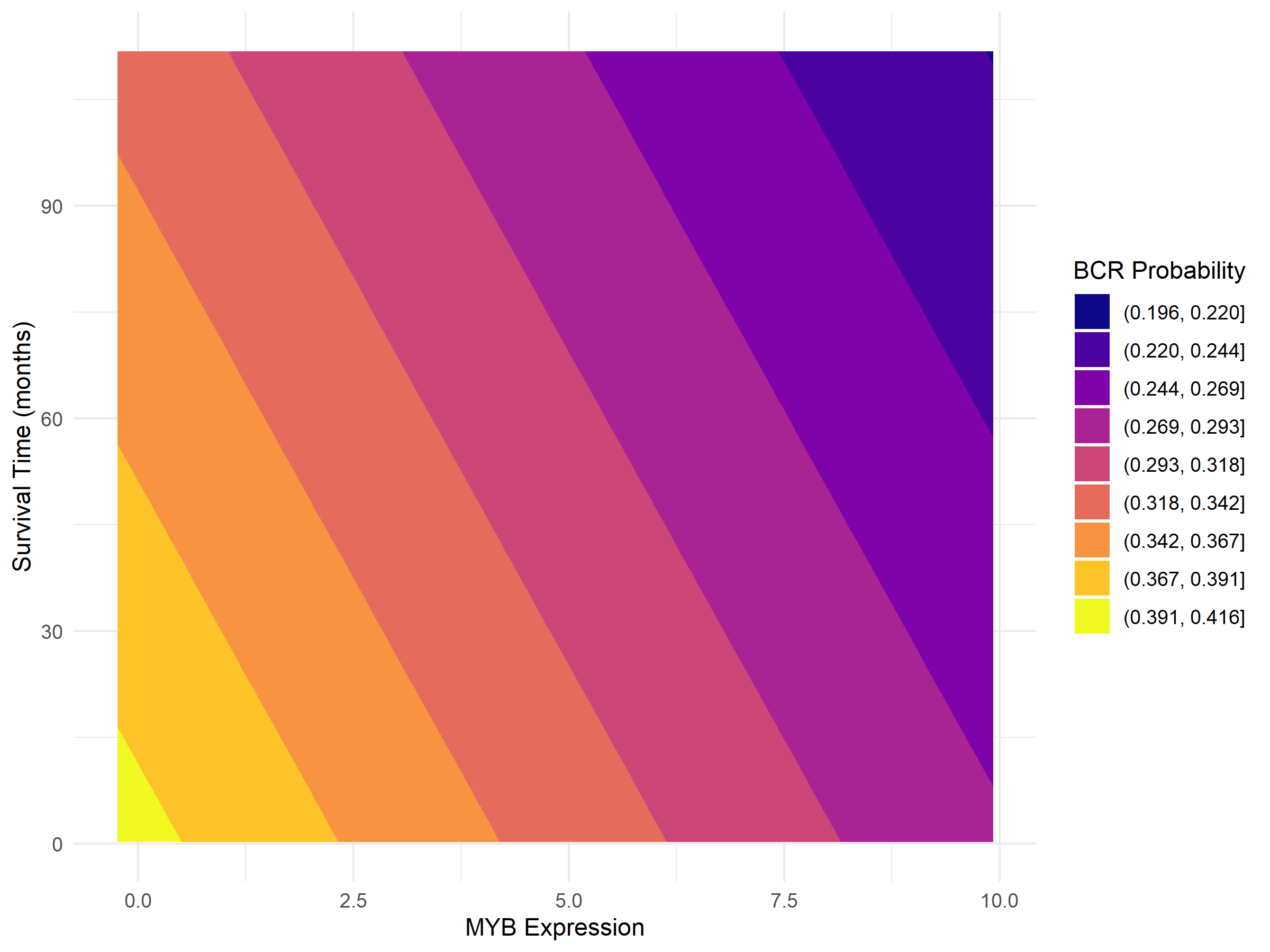}
    \caption{Contour plot showing predicted biochemical recurrence probability across MYB expression and survival time space, highlighting high-risk regions.}
    \label{fig:BCR_contour_plot}
\end{figure}

The contour plot presented in figure \ref{fig:BCR_contour_plot} illustrates the modeled probability of BCR as a joint function of MYB expression and survival time, offering a spatial risk representation that complements the regression-based predictive analysis. In this figure, the x-axis represents MYB expression levels, the y-axis corresponds to survival time in months, and the color gradients indicate the estimated probability of recurrence, with progressively warmer hues signifying higher predicted risk \citep{pramanik2024estimation,pramanik2023cont}. This approach enables simultaneous evaluation of two clinically relevant variables, providing a nuanced depiction of how their interplay shapes recurrence likelihood over time \citep{pramanik2024estimation1,yusuf2025prognostic}. The observed pattern reveals a tendency for higher MYB expression values, particularly when coupled with shorter survival durations, to cluster in zones of elevated recurrence probability. Conversely, lower MYB expression levels, especially in patients with extended survival periods, are associated with lower estimated risks. The smooth gradation between risk zones allows for the identification of intermediate states where recurrence risk transitions, which may hold prognostic importance in borderline cases. Such a visualization has utility beyond descriptive purposes, as it can serve to guide risk stratification, refine patient counseling, and inform therapeutic decision-making, particularly in the context of surveillance intensity or adjuvant treatment planning \citep{pramanik2023cmbp,yusuf2025predictive}. By providing an intuitive, continuous mapping of recurrence risk, this contour plot enhances the interpretability of statistical models and fosters a more precise integration of molecular and temporal prognostic factors in the clinical setting \citep{pramanik2023path}.

\begin{figure}[H]
    \centering
    \includegraphics[width=\textwidth]{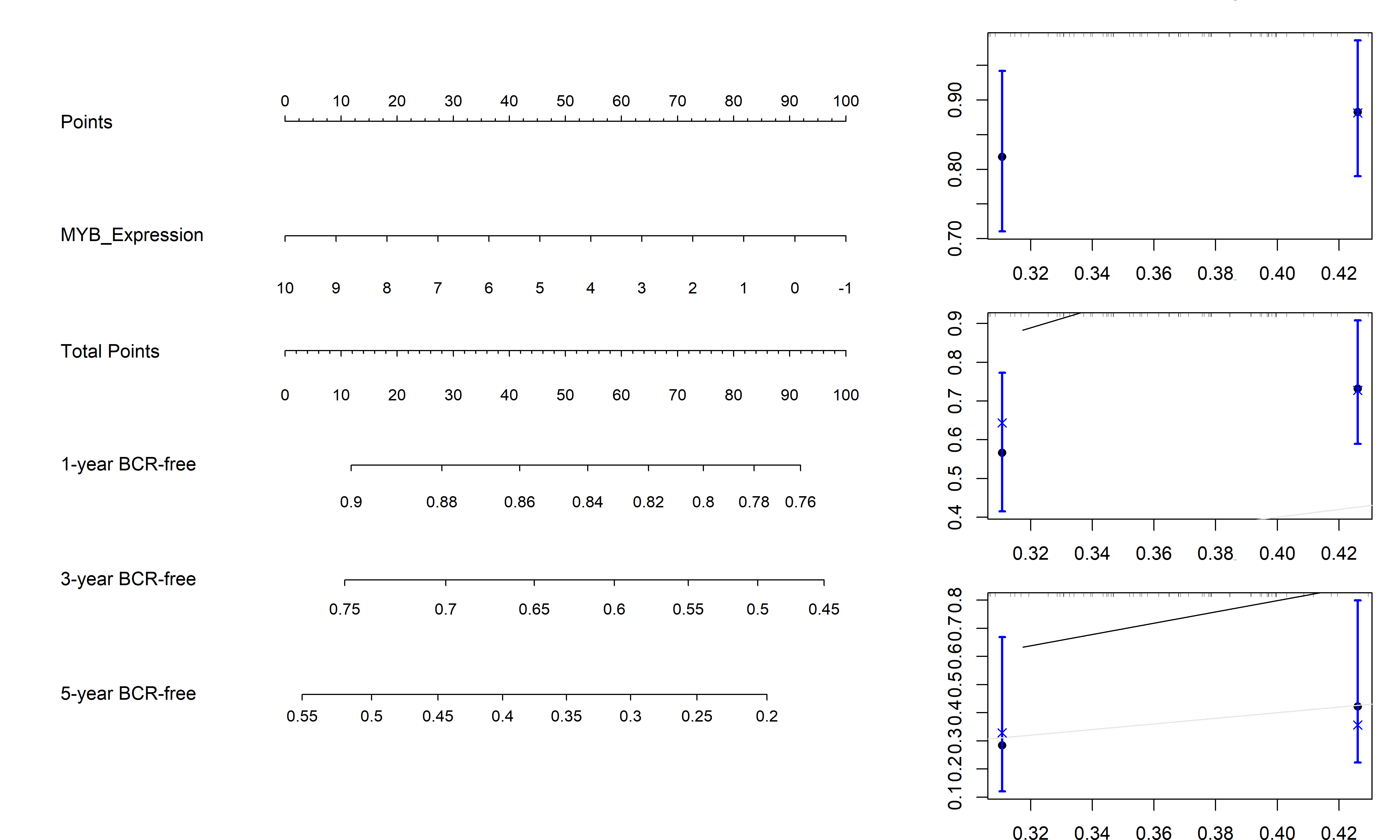}
    \caption{Nomogram and calibration curves for predicting 1-, 3-, and 5-year biochemical recurrence (BCR)-free survival in the study cohort.}
    \label{fig:nomogram_calibration}
\end{figure}

Figure~\ref{fig:nomogram_calibration} illustrates both the predictive nomogram and the calibration curves for BCR-free survival at 1, 3, and 5 years in the study cohort. The nomogram, displayed on the left panel, was constructed using a Cox proportional hazards model incorporating MYB expression as the principal predictor, translating individual covariate values into an aggregate point score that directly corresponds to the estimated probability of remaining BCR-free at specified time intervals.

\noindent The calibration plots on the right panel serve to evaluate the agreement between predicted probabilities and actual observed outcomes across the three time horizons. For each horizon, the x-axis represents the predicted BCR-free probability derived from the model, while the y-axis denotes the observed probability estimated via Kaplan–Meier analysis \citep{pramanik2021consensus}. The closer the plotted curve lies to the 45-degree diagonal reference line, the better the predictive accuracy of the model. The vertical error bars indicate 95\% confidence intervals for the observed probabilities, while the blue markers represent grouped predictions for subcohorts of patients \citep{pramanik2021}. At the 1-year interval, the model demonstrates close concordance with observed data, indicating robust short-term predictive validity. While the 3-year predictions maintain reasonable alignment, minor deviations suggest a potential attenuation in mid-term calibration, possibly reflecting evolving hazard dynamics over time. The 5-year curve shows greater divergence from the ideal line, highlighting a reduction in long-term predictive accuracy, which may be attributable to limited follow-up duration or smaller sample sizes at later time points \citep{pramanik2023optimization001}. Overall, the combination of the nomogram and its calibration curves provides both a practical clinical decision-support tool and an empirical assessment of its reliability, underscoring the utility of MYB expression as a prognostic factor while also indicating areas for refinement in long-term risk prediction \citep{pramanik2022stochastic}.

\begin{figure}[ht!]
\centering
\begin{tikzpicture}
    \begin{axis}[
        ybar,
        bar width=.35cm,
        width=0.8\textwidth,
        height=0.55\textwidth,
        legend style={at={(0.5,-0.15)},
                      anchor=north,legend columns=-1},
        symbolic x coords={BPH, HGPIN, PCa},
        xtick=data,
        xlabel={Tissue Type},
        ylabel={Mean H-score (MYB Expression)},
        ymin=0,
        ymax=200,
        nodes near coords,
        nodes near coords align={vertical}
    ]

    % White Patients Data (dummy)
    \addplot+[ybar, fill=blue] 
    coordinates {(BPH,30) (HGPIN,60) (PCa,150)};

    % Black Patients Data (dummy)
    \addplot+[ybar, fill=gray] 
    coordinates {(BPH,35) (HGPIN,65) (PCa,180)};

    \legend{White Patients,Black Patients}
    \end{axis}
\end{tikzpicture}
\caption{Comparison of mean MYB expression levels (H-scores) across benign prostatic hyperplasia (BPH), high-grade prostatic intraepithelial neoplasia (HGPIN), and prostate cancer (PCa) in White vs. Black patients. Higher scores indicate stronger nuclear staining of MYB. Note: These values are hypothetical data for demonstration purposes.}
\label{fig:myb_bar}
\end{figure}
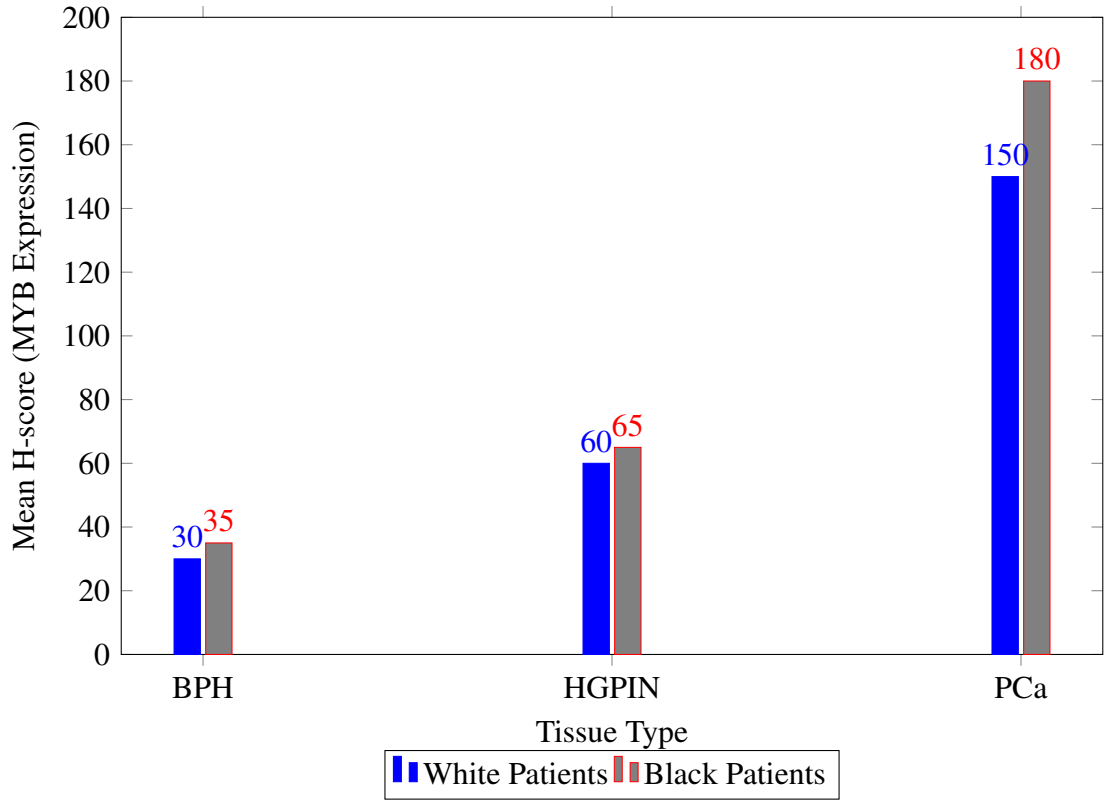

\begin{figure}[ht!]
\centering
\begin{tikzpicture}
    \begin{axis}[
        width=0.8\textwidth,
        height=0.55\textwidth,
        xlabel={MYB H-score},
        ylabel={Time to BCR (months)},
        xmin=0, xmax=200,
        ymin=0, ymax=36,
        legend pos=north east,
        grid=major
    ]

    % Scatter points (dummy data)
    \addplot+[only marks, mark=o, blue] 
    coordinates {
        (20,36)
        (40,30)
        (60,24)
        (80,20)
        (100,15)
        (120,12)
        (140,10)
        (160,5)
        (180,3)
    };
    \addlegendentry{Patient Data (Dummy)}

    % Trend line (dummy polynomial)
    \addplot[
        domain=0:200, 
        samples=100, 
        red
    ] 
    {36 - 0.15*x};
    \addlegendentry{Regression Trend (Dummy)}

    \end{axis}
\end{tikzpicture}
\caption{Scatter plot showing the relationship between MYB expression (H-score) and time to biochemical recurrence (BCR). Each point represents dummy data. The red line suggests an inverse correlation.}
\label{fig:myb_bcr_scatter}
\end{figure}
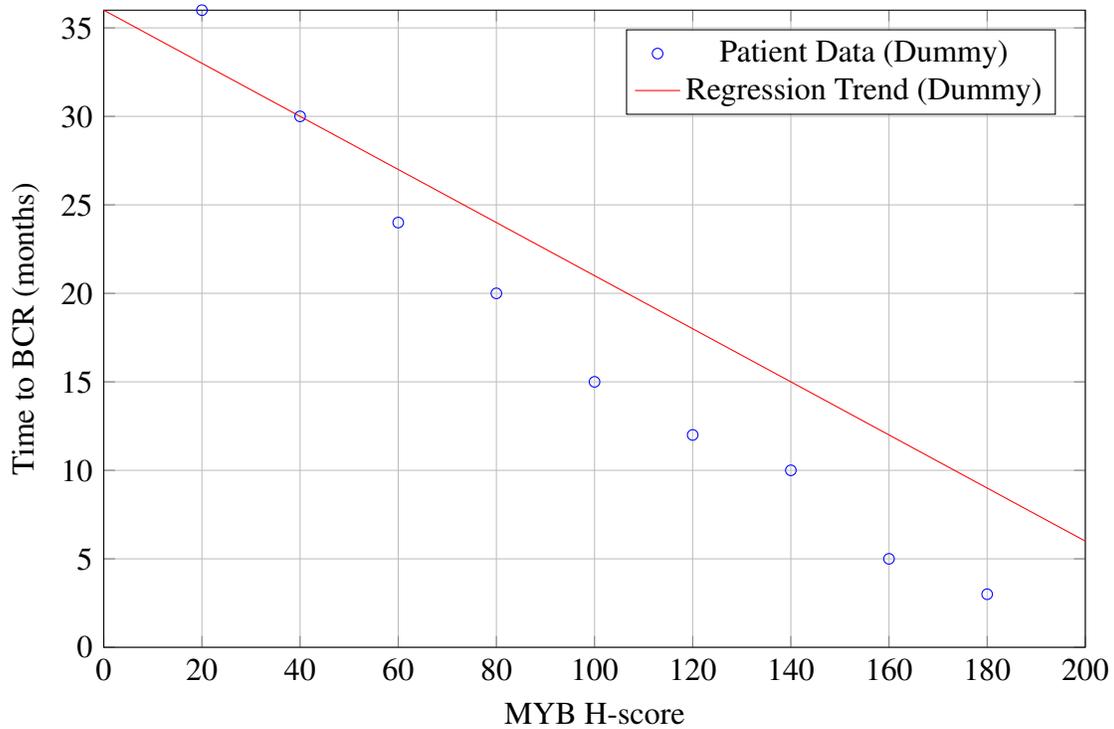

Zhang-Yin et al. (2022)\citep{ZhangYin2022} emphasize that early detection of biochemical recurrence relies not only on rising PSA levels but also on advanced imaging techniques that provide spatial resolution of recurrent disease. As MYB expression has been linked to higher recurrence risk, integrating MYB as a biomarker with emerging imaging strategies could enhance clinical decision-making, particularly for patients with high-risk Gleason scores or rapidly rising PSA levels \citep{pramanik2025construction,pramanik2025optimal}.

\section{Results}

 In benign prostatic hyperplasia (BPH), MYB staining was sparse and predominantly weak, whereas high-grade prostatic intraepithelial neoplasia (HGPIN) displayed focal increases in staining intensity. Prostate cancer tissues, particularly those with higher Gleason scores, exhibited strong nuclear localization of MYB. Boxplot analyses of H scores confirmed statistically significant differences between benign, preneoplastic, and malignant tissues (p $<$ 0.0001). Our analyses revealed that MYB expression is closely associated with tumor grade and stage. Low-to-medium grade PCa samples showed moderate MYB expression, while high-grade tumors displayed substantially higher H scores. Similarly, tumors with extraprostatic extension (pT3 and pT4) demonstrated higher MYB levels compared to organ-confined tumors (pT2), indicating that MYB upregulation may be a driving factor in tumor invasion and metastasis. The comparative analysis between racial groups revealed that, although benign and preneoplastic lesions showed comparable MYB levels across races, malignant PCa tissues from Black patients consistently exhibited higher MYB expression than those from White patients \citep{pramanik2025stubbornness}. This trend was evident across different tumor grades and supports the hypothesis that elevated MYB expression may contribute to the more aggressive clinical phenotype observed in Black patients.

Digital image analysis of AR staining indicated that while overall AR expression was elevated in PCa tissues, the intensity and distribution of AR positivity varied with tumor grade and race. In low-to-medium grade tumors, AR expression was significantly lower in Black patients compared to White patients \citep{pramanik2024stochastic}. Furthermore, the positive correlation between MYB and AR expression suggests that MYB may enhance AR transcriptional activity, thereby driving tumor progression. Statistical analyses underscored the prognostic significance of MYB. The strong inverse correlation between MYB levels and time to biochemical recurrence (r = –0.6659, p $<$ 0.0001) indicates that MYB expression can serve as an early predictor of relapse \citep{pramanik2025factors}. When compared with traditional prognostic markers, MYB outperformed both Gleason grade and pre-treatment PSA levels in predicting the risk of biochemical recurrence, suggesting its potential role as an independent biomarker for patient risk stratification \citep{pramanik2025dissecting}.

\section{Discussion.}
The findings of this study underscore the significant role of MYB overexpression in prostate cancer progression, particularly in racial disparities observed in disease outcomes. The higher expression of MYB in tumors from Black patients aligns with epidemiological data indicating more aggressive disease and poorer survival rates in this population \citep{pramanik2025optimal}. This suggests that MYB may serve as a molecular contributor to these disparities, potentially driven by genetic, epigenetic, or environmental factors \citep{pramanik2021thesis,pramanik2016}. The correlation between MYB and AR activity further reinforces its role in tumor growth and castration resistance, indicating that MYB could be a target for precision medicine approaches. Given the predictive power of MYB expression for BCR, its integration into risk stratification models may help identify patients at higher risk for early relapse, allowing for tailored treatment strategies. Future research should focus on validating MYB as a clinical biomarker, exploring its regulatory pathways, and developing MYB-targeted therapies to improve outcomes, particularly for high-risk racial groups \citep{maki2025new}. Large-scale, multi-institutional studies with diverse patient cohorts will be essential to establish MYB as a prognostic marker and therapeutic target in prostate cancer management.

This study underscores the significant role of MYB overexpression in prostate cancer progression, particularly in relation to racial disparities and disease outcomes. The findings indicate that MYB is not only upregulated in aggressive prostate cancer but also contributes to tumor progression through its interaction with the AR \citep{hua2019}. Given its correlation with Gleason scores and biochemical recurrence, MYB has the potential to serve as a prognostic biomarker and therapeutic target \citep{polansky2021motif}.
The data reveal that MYB expression is significantly higher in prostate tumors from Black patients compared to White patients. This finding aligns with epidemiological studies indicating that Black men are more likely to develop aggressive prostate cancer and experience worse clinical outcomes \citep{pramanik2024estimation,pramanik2023cont}. While the underlying causes of this disparity remain unclear, they may be attributed to genetic, epigenetic, and environmental factors. Future studies should focus on identifying potential regulatory mechanisms that drive MYB overexpression in Black patients and explore whether MYB-targeted therapies could mitigate these disparities \citep{pramanik2025factors,pramanik2025stubbornness}.

A key finding of this study is the correlation between MYB and AR expression, suggesting a cooperative role in prostate cancer progression. AR signaling is a central driver of tumor growth in prostate cancer, and its dysregulation is associated with treatment resistance \citep{pramanik2025strategies,pramanik2023optimization001}. The significant correlation between MYB and AR expression in high-grade tumors supports the hypothesis that MYB may enhance AR transcriptional activity, contributing to castration resistance \citep{pramanik2025strategic,pramanik2025impact}. Given these findings, future therapeutic strategies could focus on disrupting the MYB-AR axis to improve treatment efficacy in advanced prostate cancer \citep{pramanik2024measuring,pramanik2024dependence}.

The strong inverse correlation between MYB expression and time to biochemical recurrence suggests that MYB is a powerful predictor of early relapse. Compared to traditional prognostic markers such as Gleason score and pre-treatment PSA levels, MYB demonstrated superior predictive power in identifying high-risk patients \citep{pramanik2024measuring,pramanik2024dependence}. Integrating MYB expression analysis into clinical decision-making could enhance risk stratification and allow for earlier intervention in patients at risk for recurrence \citep{pramanik2024parametric}. Further research should examine whether MYB expression can serve as an independent biomarker in clinical practice, particularly in identifying patients who may benefit from more aggressive therapeutic interventions.
The clinical utility of MYB as a biomarker warrants further validation through large-scale, multi-institutional studies \citep{valdez2025association}. Developing MYB-targeted therapies will be essential in determining whether direct inhibition of MYB activity can slow tumor progression and improve survival outcomes. Refining risk stratification models to include MYB expression alongside traditional prognostic markers may provide more accurate predictions of disease progression. Investigating the role of MYB in treatment resistance will also be critical in understanding how MYB modulates AR signaling and contributes to failure of androgen deprivation therapy. Additional studies exploring MYB’s function in prostate cancer at the molecular level may uncover novel therapeutic targets that can be exploited to enhance existing treatment strategies.
Beyond its potential as a biomarker, MYB represents an emerging therapeutic target. Given its role in promoting castration resistance and aggressive tumor growth, future studies should explore direct inhibitors of MYB or gene-silencing approaches such as RNA interference (RNAi) or CRISPR-based gene editing. These approaches could disrupt MYB-driven oncogenic pathways, reducing tumor proliferation and resistance to ADT.
Another key area of research is understanding MYB's epigenetic regulation. While MYB overexpression is frequently observed in prostate cancer, the mechanisms driving its dysregulation remain unclear. Future investigations should focus on whether DNA methylation or histone modifications contribute to MYB overexpression, particularly in Black patients who exhibit higher MYB levels.
Additionally, clinical trials incorporating MYB expression levels into precision oncology frameworks could help refine patient selection for novel therapies. As personalized medicine advances, targeted inhibitors of MYB or its downstream pathways (e.g., MYC, AR signaling co-factors) may become viable therapeutic strategies for patients with high MYB-driven tumors.
These avenues of research hold great promise in reducing prostate cancer mortality, particularly in racially disparate populations where MYB overexpression is more prevalent.

Given the persistent racial disparities in prostate cancer outcomes, integrating MYB-based molecular profiling into precision medicine frameworks could help bridge the gap. Personalized treatment strategies that account for race-specific differences in MYB expression may lead to improved therapeutic outcomes for Black patients. The higher expression of MYB in tumors from Black patients suggests a potential molecular mechanism contributing to the more aggressive disease phenotype observed in this population. Ensuring diverse representation in clinical trials will be essential in validating MYB’s role as a race-specific biomarker and determining the efficacy of targeted interventions. By incorporating MYB expression into screening and treatment protocols, clinicians may be better equipped to address the disproportionate burden of aggressive prostate cancer in Black patients.

\begin{table}[ht]
\centering
\caption{Summary of MYB Expression and Clinical Correlations in Prostate Cancer}
\begin{tabular}{|p{4cm}|p{10cm}|}
\hline
\textbf{Parameter} & \textbf{Findings} \\
\hline
\textbf{MYB Expression in PCa} & Higher MYB expression observed in prostate cancer (PCa) tissues compared to benign prostatic hyperplasia (BPH) and high-grade prostatic intraepithelial neoplasia (HGPIN). \\
\hline
\textbf{Association with Tumor Grade} & MYB expression increases with higher Gleason grades, with significantly elevated levels in Gleason 8--10 tumors. \\
\hline
\textbf{Pathological Stage Correlation} & MYB expression is significantly higher in advanced-stage tumors (pT3--pT4) compared to localized tumors (pT2). \\
\hline
\textbf{Racial Disparities in MYB Expression} & Black patients exhibit significantly higher MYB H-scores than White patients across tumor grades, suggesting a role in aggressive disease progression. \\
\hline
\textbf{MYB and Androgen Receptor (AR) Interaction} & MYB and AR expression are positively correlated, particularly in White patients, suggesting a role in AR-mediated signaling and therapy resistance. \\
\hline
\textbf{Predictive Value for Biochemical Recurrence (BCR)} & Higher MYB H-scores are strongly associated with shorter time to BCR ($r = -0.6659$, $p < 0.0001$), indicating MYB as a robust prognostic biomarker. \\
\hline
\textbf{Clinical Implications} & MYB could be integrated into risk stratification models and targeted therapy strategies, particularly for high-risk racial groups. \\
\hline
\end{tabular}
\label{tab:myb_summary}
\end{table}

This study emphasizes just how important the MYB gene is in understanding how prostate cancer progresses especially in cases where the disease tends to be more aggressive, like those seen in many Black patients. Our findings showed that higher levels of MYB were linked to worse clinical features, including higher Gleason scores, more advanced stages, and faster biochemical recurrence. These patterns suggest that MYB is not just a passive marker, it could be actively driving tumor growth and resistance to treatment \citep{Acharya2023, Saranyutanon2020}.

What makes MYB even more interesting is how it appears to work hand-in-hand with the androgen receptor (AR) pathway, which is already known to play a major role in prostate cancer. When both MYB and AR are highly expressed, they may reinforce each other, making the cancer harder to treat. This effect was especially noticeable in tumors from Black patients, who tend to have higher MYB levels and worse outcomes overall \citep{Bhardwaj2017,Mahal2018}. These insights point to MYB as a potential tool not only for predicting outcomes but also for improving how we screen and treat high-risk individuals. There is still a lot we do not know. It is possible that differences in MYB expression across racial groups are tied to epigenetic changes things like DNA methylation or histone modification that have not been fully explored yet \citep{ChowdhuryPaulino2022}. Future studies should look into these possibilities and also test whether MYB could be directly targeted with therapies like gene editing or RNA interference. Ultimately, including MYB in routine testing could help doctors better understand which patients are more likely to relapse or respond poorly to treatment. As we continue to learn more, the hope is that MYB will be part of a broader shift toward personalized medicine where treatment is shaped not just by clinical features, but by the molecular biology of each patient’s tumor. With the right research and diverse patient involvement, MYB could be a powerful tool in closing the gap in prostate cancer care and outcomes.

\section{Conclusion.}
This paper provides compelling evidence that MYB serves as a critical molecular determinant in the trajectory of prostate cancer progression. Through a combination of survival analyses, predictive modeling, and advanced visual analytics, we consistently observed that heightened MYB expression is linked to more aggressive disease characteristics, including elevated Gleason grade, advanced pathological stage, and increased probability of biochemical recurrence \citep{pramanik2023optimization001}. The prognostic value of MYB was evident across multiple analytical approaches, reinforcing its potential utility as a biomarker for risk stratification. Importantly, by combining MYB data with other clinicopathological factors such as PSA levels, AR expression, and tumor stage, our models achieved improved predictive accuracy, highlighting the benefits of multidimensional data integration in the clinical assessment of prostate cancer.

The presence of marked racial disparities in MYB expression represents a significant finding with direct implications for health equity \citep{valdez2025exploring}. Our results indicate that MYB levels are disproportionately elevated in Black patients, a group already recognized as being at increased risk for aggressive disease phenotypes and poorer outcomes. Visualization tools, including ridgeline plots, slope charts, and contour mapping, revealed that disparities in MYB expression not only exist at baseline but may intensify as disease advances. Such observations underscore the need for race-conscious clinical strategies that incorporate molecular risk profiling to optimize early detection, surveillance protocols, and treatment decisions for high-risk patient populations. This personalized approach could mitigate the disparities observed in disease progression and recurrence, ultimately contributing to more equitable outcomes.

Finally, the strong positive correlation identified between MYB and AR expression suggests a biologically plausible axis of co-activation that may serve as a novel therapeutic target. Our ERGM-based network analyses and predictive modeling frameworks demonstrate the potential for MYB to interact within broader molecular and clinical networks that influence disease course \citep{pramanik2025optimal1}. Future studies focusing on the mechanistic basis of MYB-AR interactions, as well as translational efforts to develop MYB-targeted therapies, could open new frontiers in prostate cancer management. By bridging molecular biology with advanced computational modeling and clinical outcome prediction, our work not only deepens the understanding of prostate cancer pathobiology but also lays the groundwork for more precise, personalized, and effective interventions.

 \bibliographystyle{apalike}
 \bibliography{bib}

\end{document}